\documentclass[acmtog]{acmart}

\usepackage{booktabs} 

\DeclareMathOperator*{\argmin}{argmin}

\newcommand{\norm}[1]{\left\lVert#1\right\rVert}

\definecolor{greend}{rgb}{0.0, 0.5, 0.0}

\def \NVP {\textit{Neural Voice Puppetry}}

\def \etal {et al.}

\usepackage[ruled]{algorithm2e} 

\SetAlFnt{\small}
\SetAlCapFnt{\small}
\SetAlCapNameFnt{\small}
\SetAlCapHSkip{0pt}

\acmJournal{TOG}
\acmVolume{39}
\acmNumber{6}
\acmArticle{268}
\acmYear{2020}
\acmMonth{12}
\acmDOI{10.1145/3414685.3417808}

\setcopyright{rightsretained}

\begin{document}
\title{Egocentric Videoconferencing}

\author{Mohamed Elgharib}
\authornote{Indicates equal contribution}
\authornote{Contact us through: elgharib@mpi-inf.mpg.de}
\email{elgharib@mpi-inf.mpg.de}
\affiliation{%
  \institution{Max Planck Institute for Informatics, SIC}}
\author{Mohit Mendiratta}
\authornotemark[1]
\email{mmendira@mpi-inf.mpg.de}
\affiliation{%
  \institution{Max Planck Institute for Informatics, SIC}}
\author{Justus Thies} 
\email{justus.thies@tum.de}
\author{Matthias Nießner}
\email{niessner@tum.de}
\affiliation{%
  \institution{Technical University of Munich}}
\author{Hans-Peter Seidel}
\email{hpseidel@mpi-sb.mpg.de}
\author{Ayush Tewari}
\email{atewari@mpi-inf.mpg.de}
\affiliation{%
  \institution{Max Planck Institute for Informatics, SIC}}
\author{Vladislav Golyanik}
\email{golyanik@mpi-inf.mpg.de}
\author{Christian Theobalt}
\email{theobalt@mpi-inf.mpg.de}
\affiliation{%
  \institution{Max Planck Institute for Informatics, SIC}}

\renewcommand\shortauthors{Elgharib, M. and Mendiratta, M. et al.} 

\begin{CCSXML}
<ccs2012>
<concept>
<concept_id>10010147.10010371.10010382</concept_id>
<concept_desc>Computing methodologies~Image manipulation</concept_desc>
<concept_significance>500</concept_significance>
</concept>
</ccs2012>
\end{CCSXML}

\ccsdesc[500]{Computing methodologies~Computer graphics}
\ccsdesc[300]{Computing methodologies~Image manipulation}
\ccsdesc[300]{Computing methodologies~Animation}
\ccsdesc[300]{Computing methodologies~Rendering}

\keywords{Videconferencing, Egocentric, Face Frontalisation, Neural Rendering, Reenactment, Face.} 

\begin{abstract}

We introduce a method for egocentric
videoconferencing 
that enables hands-free video calls, for instance %
by people wearing smart glasses or other mixed-reality devices.
Videoconferencing portrays valuable non-verbal communication and face expression cues, but usually requires a front-facing camera.
Using a frontal camera in a hands-free setting when a person is on the move is impractical.
Even holding a mobile phone camera in the front of the face while sitting for a long duration is not convenient. 
To overcome these issues, we propose a low-cost wearable egocentric camera setup that can be integrated into smart glasses.
Our goal is to mimic a classical video call, and therefore, we transform the egocentric perspective of this camera into a front facing video.
To this end, we employ a conditional generative adversarial neural network that learns a transition from the highly distorted egocentric views to frontal views common in videoconferencing. 
Our approach learns to transfer expression details directly from the egocentric view without using a complex intermediate parametric expressions model, as it is used by related face reenactment methods.
We successfully handle subtle expressions, not easily captured by parametric blendshape-based solutions, e.g., tongue movement, eye movements, eye blinking, strong expressions and depth varying movements. 
To get control over the rigid head movements in the target view, we condition the generator on synthetic renderings of a moving neutral face.
This allows us to synthesis results at different head poses. 
Our technique produces temporally smooth video-realistic renderings in real-time using a video-to-video translation network in conjunction with a temporal discriminator.
We demonstrate the improved capabilities of our technique by comparing against related state-of-the art approaches. 
\end{abstract}

\begin{teaserfigure}
   \centering
   \includegraphics[width=1.0\textwidth]{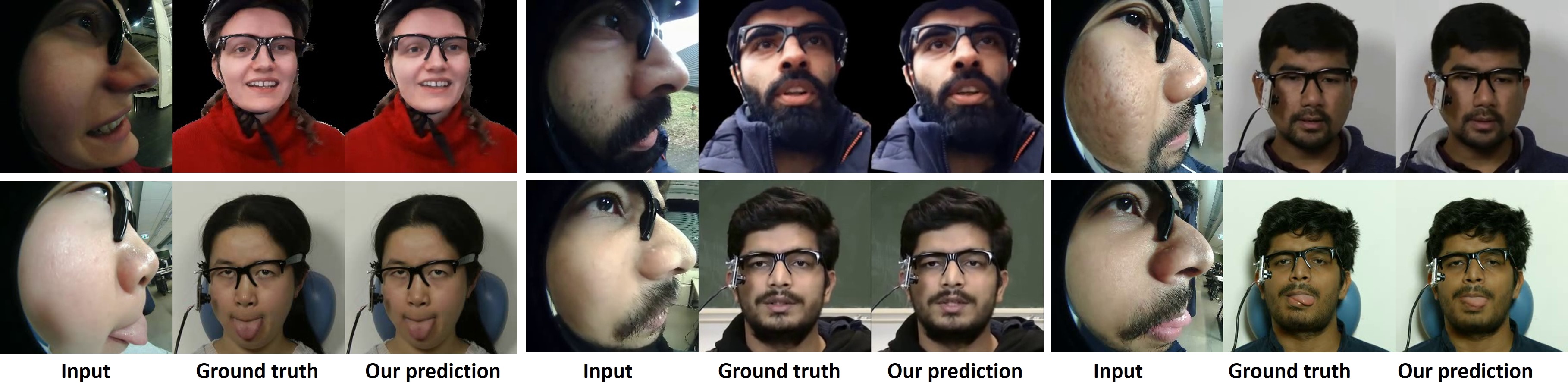}
   \caption{We present an approach for hands-free videoconferencing. Given the view of an egocentric camera, that is attached to an eye-glasses frame, we predict a frontalised video stream which is common in videoconferencing. 
   }
   \label{fig:teaser}
\end{teaserfigure}

\maketitle

\section{Introduction}

Videoconferencing is popular as it portrays a wide range of communication signals, beyond traditional phone calls that are used, to face-to-face conversations that use visual cues such as facial expressions or eye gaze. 
This improves engagement in conversations.  
Video calls usually require a camera observing the face from a frontal perspective to allow good facial coverage.
While feasible in controlled and static indoor settings, \textit{e.g.,} when working at your desk, such camera placement is not feasible in many other everyday scenarios where people call each other with mobile devices, especially, when walking outdoors in dynamic environments.
In such outdoor settings, or even when walking around or just sitting at home, holding up a camera or mobile phone in front of your face for a long duration to transmit a frontal video of yourself is not viable.   

Recently, reenactment algorithms are combining model-based and learning-based scene reconstruction and rendering. They showed the ability to control and modify facial expressions and poses in video filmed from a frontal perspective in a highly photo-realistic manner \cite{kim2018DeepVideo,pagan,thies2019,thies2019nvp,Suwajanakorn2017}.
This enables a new level of visual quality in various applications, such as avatar generation \cite{pinscreen}, visual dubbing \cite{Kim19NeuralDubbing}, video post-production \cite{Fried2019}, and virtual assistant generation \cite{thies2019nvp}.
Similar methods were also used to improve virtual reality teleconferencing with head-mounted displays (HMDs). 
Here, it was shown that reenactment techniques, in combination with additional sensors placed inside and outside of the HMD, enable it to virtually remove the display from a user's face, and thus portray an unobstructed view of the user or her avatar to every participant of the VR teleconference~\cite{Lombardi2018,Lombardi2019,olszewski2016high,thies2018facevr}. 
However, wearing such non-see-through VR HMDs while walking around in general environments is impractical. 
We, therefore, present a new approach to enable people to start hands-free video calls even when they roam around in general indoor and outdoor environments. 
We want our approach to be similarly convenient and non-encumbering as audio calls with a headset, while additionally transmitting a frontal video image of the person.
To this end, we propose an egocentrically worn hardware setup and a new algorithm to achieve this goal. 
Our approach uses a single commodity RGB fish-eye camera mounted to the side of an eye glass frame. 
While our prototype setup currently has a larger form factor, we argue that mass production of future smart and augmented reality glasses will make it easy to integrate starkly miniaturised cameras in this way.
The related designs of Google Glasses or the Snapchat Spectacles have shown this.
While a camera placed in this way minimizes obstruction to the user's field of view, it records a highly distorted and incomplete facial view that is not directly suitable for viewing in a videoconferencing application (see Fig.~\ref{fig:teaser}, input).
We therefore present a new conditional generative adversarial network that learns to frontalise the full face in real-time given the starkly distorted egocentric view as input.
An adversarial loss that operates on a sequence of estimated frontalised images ensures temporal consistency while a perceptual loss is employed to produce high fidelity results.
Our frontalisation algorithm purposefully refrains from estimating an intermediate representation of the full face performance on the basis of a 3D morphable face model (3DMM), as it was done in many previous face reenactment methods~\cite{kim2018DeepVideo,thies2019}. 
3DMMs usually lack the variable dimensions to represent all fine-grained nuances of eye gaze, eye blinks, facial micro-expressions, or expressive mouth and tongue motion, which are important non-verbal cues in face-to-face communication. 
In addition, even if these dimensions were parametrically modeled, estimating them from a starkly oblique and distorted view is non-trivial.
Therefore, our frontalisation method only uses weak conditioning with a neutral 3D face model without face expressions, and transfers the fine-grained expression details from the egocentric view to the frontal view by means of learned direct video-to-video mapping.  

Our lateral fisheye-to-frontal transfer method solves a much harder problem than established frontalisation settings  \cite{Yin_2017_ICCV,Peng_2017_ICCV,Zhang_2019,Cao2019}, and unlike these it produces temporally coherent and photo-realistic renderings with good audio-lip sync. 
Our technique photo-realistically captures and frontalises a wide range of important expression details, eye gaze and eye blinking, for which parametric expression model-based solutions would struggle \cite{kim2018DeepVideo,thies2019,thies2019nvp}.
Our approach captures the lighting of a person's  surroundings by observing the egocentric view and reproduces it in the frontal view.
We also demonstrate that adapting purely audio-driven methods for face  reenactment~\cite{thies2019nvp,Suwajanakorn2017} to our frontalisation task does not suffice since subtle facial expression cues are not uniquely correlated to speech and yet clearly appear in the egocentric video. 
Our solution is trained in a supervised manner, without manual annotations. 

\vspace{\baselineskip}
\noindent
To summarise, we make the following contributions: 
\begin{itemize}
    \item A light-weight capturing setup that enables hands-free videoconferencing and is easy to be integrated into smart glasses.
    \item A real-time video-to-video translation technique that uses a new conditional neural network adversarially trained with a temporal discriminator to transfer even subtle face expression details and extreme face expressions from an egocentric fisheye to a frontal view.
    \item A thorough analysis showing that our approach reconstructs frontal video $54\%$ more accurately than established image-to-image translation methods (pix2pix~\cite{IsolaZZE2017}), visually outperforms 3DMM expression based solutions, and runs in real time at 29.4 ms per frame.
\end{itemize}

\section{Related Work}

We survey computer graphics and vision techniques that can potentially enable videoconferencing in our scenario of a an egocentric input view. 
Current techniques can be classified into frontalisation-based and reenactment-based approaches.
Frontalisation techniques produce a frontal complete view of the face from an incomplete side view, while reenactment methods transfer facial motions onto prerecorded video~\cite{thies2016face,kim2018DeepVideo,thies2019nvp}.
Unlike face frontalisation, reenactment focuses more on photo-realistic editing in the same camera perspective.

\subsection{Face Frontalisation} 

Face frontalisation techniques 
are commonly designed to transform large or profile face poses in a camera view, where larger parts of the face are occluded, into complete and frontal views of the face.
Existing approaches can be divided into face model based \cite{Hassner_2015_CVPR,Zhu16,Peng_2017_ICCV,Yin_2017_ICCV,Cao_NIPS2018,Cao2019} and image-to-image translation-based \cite{Sagonas_2015,Zhang_2019,Wiles18,IsolaZZE2017}. Model based techniques use a parametric 3D Morphable Model (3DMM) to represent faces \cite{garrido2016reconstruction}. Such model provides a parametric control over the head pose and hence allows frontalising the face as observed from a front looking camera. 
Zhu~\etal~\cite{Zhu16} and Peng~\etal~\cite{Peng_2017_ICCV} learn the parameters of the face model from the input face using a deep neural network. The network is trained using pairs of profile and frontal view faces. Such data is obtained by synthesising the profile images from the frontal images. 
In Peng~\etal~\cite{Peng_2017_ICCV}, the frontalisation network also learns to disentangle the identity from the head pose. 
Yin~\etal~\cite{Yin_2017_ICCV} use a GAN conditioned on the the frontalised synthetic rendering to produce photo-realistic results. The output is also constrained to maintain the identity of the examined image. 
The approach of Cao~\etal~\cite{Cao_NIPS2018,Cao2019} learns to frontalise the input by accessing the face texture through a uv-map. 
A discriminator loss learns to differentiate between frontal and non-frontal views.  

Image-translation based techniques 
\cite{Sagonas_2015,Zhang_2019,Wiles18} learn to frontalise faces without a parametric 3D model. 
This bypasses some limitations imposed by the restricted expressiveness of learned parametric expression and shape models.
Sagonas~\etal~\cite{Sagonas_2015} observe that frontal faces have the minimum rank across all poses. They formulate face frontalisation as an optimisation problem with nuclear norm minimisation. A statistical prior is learned  from frontal images. Zhang~\etal~\cite{Zhang_2019} propose a flow based approach for face frontalisation. A flow field is initialised by SIFT-flow \cite{SIFTFLOW} and refined through a convolutional neural network. 
Most current frontalisation technique focus more on improving facial recognition techniques and not on producing photo-realistic and temporally coherent video outputs \cite{Yin_2017_ICCV,Cao_NIPS2018, Cao2019,Sagonas_2015,Zhang_2019}. 
In contrast, our method produces photo-realistic frontalisations that are temporally coherent. It does so by translating between a strongly distorted fish-eye camera and a more regular frontal camera (see Fig.~\ref{fig:teaser}).

Paired image-to-image translation techniques \cite{Wiles18,IsolaZZE2017} can also be used for frontalisation. X2face \cite{Wiles18} train a network to extract a face embedding based on a single image. The extracted latent code is used to synthesise a new image of the target face with new expressions. Isola ~\etal's pix2pix \cite{IsolaZZE2017} uses a Conditional Generative Adversarial Network (CGAN) to translate an input image from one domain to another. An adversarial loss pushes the output to resemble the ground truth. 
While pix2pix shows interesting results, it is not specifically designed for faces and hence has no prior knowledge neither on face structure nor movement. 
It also processes each frame in isolation. 
We demonstrate that applying pix2pix to our use-case generates noticeable artifacts and deformations in the face structure. 

\subsection{Reenactment-Driven Solutions} 
 
Facial reenactment is the process of capturing the face expression and pose from a source actor in video and transferring them to video of a different target face.
Many recent reenactment approaches rely on model-based expression capturing.
In contrast to classical computer graphics approaches that render the modified target face on top of the input video using a static face texture~\cite{thies2016face,Garrido2014} or a dynamic texture~\cite{thies2018headon}, neural rendering approaches replace components of the standard graphics pipeline by learned components.
Deep Video Portraits~\cite{kim2018DeepVideo} proposes an image-to-image translation approach that converts synthetic renderings to realistic imagery.
This approach is inspired by pix2pix~\cite{IsolaZZE2017} and uses a U-Net architecture as a generator that gets synthetic renderings of the underlying 3DMM as input.
Kim~\etal~\cite{Kim19NeuralDubbing}
presented a reenactment technique which maintains the speaking style of the target identity. The work shows the importance of this feature during visual dubbing. paGAN~\cite{pagan} generates a personalised avatar from a single image. A translation network trained on several identities learns to bridge the gap between a model-based rendering and its corresponding photo-real version. 
Deferred Neural Rendering~\cite{thies2019} introduced neural textures which store high-dimensional neural descriptors. Such textures are interpreted by a neural network to produce a photo-realistic output.
Both, neural texture and the interpreting network are trained in conjunction based on a short target video sequence, where the 3DMM face model parameters are used to render the neural texture to image space.
While the approach is general and can be applied to novel view point synthesis and scene editing, they also demonstrate high quality facial re-rendering and reenactment.
In contrast to the model-based approaches, there are also techniques that are not relying on a 3DMM face model prior. Based on a sophisticated multi-camera setup, Lombardi~\etal~\cite{Lombardi2018} learn a deep appearance model, that takes a image as input to predict a latent model descriptor that is interpreted by a decoder network. In a follow-up work they show the ability of driving a highly photo-realistic avatar through a VR-headset~\cite{Wei2019VRFA}.

Recently audio and text driven reenactment approaches have been proposed. Here facial expressions are not extracted from a source video, but rather from a source audio or a text script. Fried~\etal's~\cite{Fried2019} text-based editing technique maps phonemes to the expression parameters of the 3DMM. This allows text-based reenactment in photo-realistic and temporally smooth manner. Several approaches for audio-driven reenactment are available 
\cite{Suwajanakorn2017,thies2019nvp,Chung17b,olszewski2016high}. 
Thies~\etal~\cite{thies2019nvp} presented \NVP, an approach for estimating facial expressions from the audio and rendering it in a photo-realistic manner. 
They use DeepSpeech~\cite{Hannun2014DeepSS} to produce character logits from the input audio. 
A network then translates the logits into the parameters of a blendshape expression model. A synthetic rendering of a face model is produced, followed by neural rendering for photo-realistic results. 
Swajanakorn~\etal~\cite{Suwajanakorn2017} define the mouth shape with a number of keypoints, for which they regress their positions from only the audio signal. 
With proper image compositing of the predicted mouth shapes, they show impressive high quality visual renderings of the 44th president of the United States Barack Obama. 
Olszewski et al. \cite{olszewski2016high} used the audio signal to assist performance capture from a VR-headset. Here, audio-based alignment techniques are used to map same utterance from different subjects into the same animation parameters. 
We show that purely audio-driven solutions do not suffice in our egocentric videoconferencing setting since important non-verbal expressions only appear on video. 

\section{Data Recording}
\label{sec:data}

\begin{figure*}[h]
	\centering
	\includegraphics[width=0.9\linewidth]{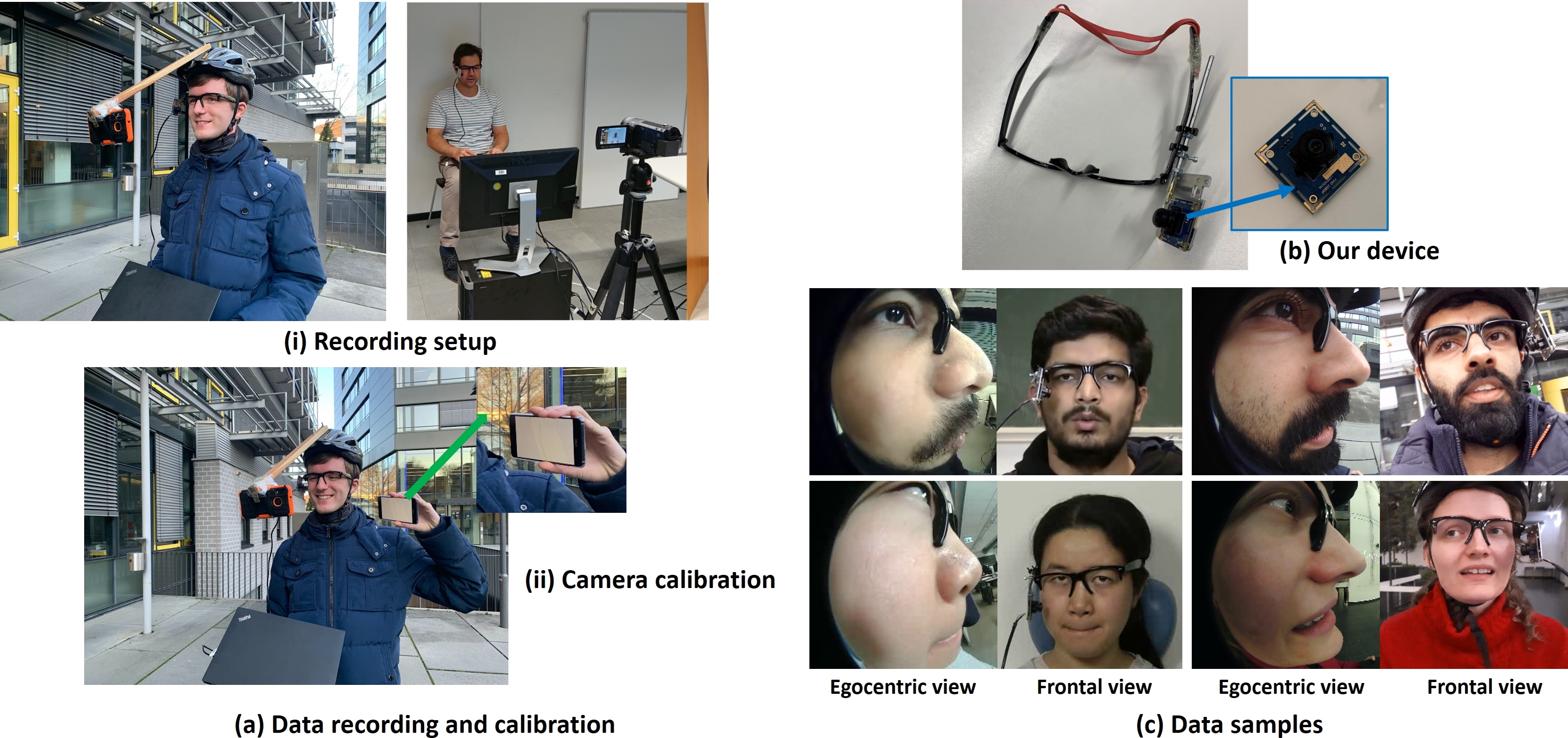}
	\caption
	{
	    (a) Proposed low-cost data capturing setup. 
		We show the setup for the dynamic ((i)-left) and sitting ((i)-right) scenarios. The egocentric and front cameras are synchronised by observing a simple transient event ((ii)-zoom on region).
		(b) We use a commodity RGB fish-eye camera attached to the side of an eye glass frame. This captures the face from an egocentric perspective, which is passed to our frontalisation approach. {Our device placement is always tightened by a rubber band around the head to limit the camera movements/shakiness during recordings.}
		(c) Samples of data generated by our recording setups. Our approach learns to translate the egocentric view into the frontal view.
		}
	\label{fig:recordingdevicedata}
\end{figure*}

We propose a recording setup for obtaining the training data of our solution
(see Fig.~\ref{fig:recordingdevicedata}-a).
The training data consists of paired egocentric and front-view videos recorded using commodity RGB cameras.
The recorded videos are temporally synchronised using a simple calibration stage. 
We have two data recording setups: one for a dynamic scenario and another for a sitting scenario (see Fig.~\ref{fig:recordingdevicedata}-a, top row). They both use the same egocentric camera but differ in the setup of the frontal camera. 

{Egocentric Camera:}
We use a low-cost RGB fish-eye camera to capture the facial expressions (ELP-usbfhd01m-l180).
The camera is attached to the frame of an eye glass such that it minimizes obstruction to the user field of view and maximizes the face coverage (see Fig.~\ref{fig:recordingdevicedata}-b).
It has a diagonal field of view of 180 degrees, and records images with resolutions up to $1280 \times 1960$ at 30fps. %
Fig.~\ref{fig:recordingdevicedata}-c shows data samples captured by the egocentric camera. 
During test time, these views are the input to our method.
Specifically, our algorithm learns to estimate the corresponding full face frontalisation. This learning is supervised by a frontal camera.  

{Frontal Camera:}
We use a monocular RGB frontal camera placed in front of the user to capture the face from the target perspective.
All our experiments in the dynamic scenario are shot using a commodity mobile phone camera. A commodity HD camera is used for recording our experiments in the sitting scenario.
The dynamic scenario resembles situations where the user is moving around in an environment with changing illumination and background. 
In this case, our supervising camera is attached to a regular bicycle helmet in a way to allow good face coverage (see Fig.~\ref{fig:recordingdevicedata}-a, top-left), capturing the face from a frontal view at a fixed location with respect to the face.
In the sitting scenario we place the supervising camera on a tripod, and the user sits in front of it (see Fig.~\ref{fig:recordingdevicedata}-a top-right).
In many sequences the user reads a collection of 111 english pangrams\footnote{\url{https://callibeth.com/downloads/pangrams111.pdf}} while being recorded. 
The pangrams are read from a laptop screen.
Each pangram is a sentence containing all the 26 Latin letters. This captures a wide variety of visemes commonly used in everyday speeches. 
In other sequences, subjects were asked to talk freely, imitating a phone call,  discussing a popular topic and so on.
In the dynamic scenario the user walks, in either outdoor or indoor environment. 
In the sitting scenario the user sits on a chair and moves his/her head naturally and freely. 
To synchronise the egocentric and front camera recordings, we use a transient event.
\textit{i.e.,} a mobile phone screen observed from both cameras plays a video of mostly black frames, but with a single white frame every $10$ seconds (see Fig.~\ref{fig:recordingdevicedata}-a, bottom, zoom on region).
We start recording from both cameras and wait until the white frame is observed.
We use this white frame to temporally synchronise the egocentric and frontal recordings. 
The calibration is only done once at the start of the data recording. 
We recorded 27 sequences with 13 identities, and extracted them at 24 frames per second. Sequences are on average $14000$ frames long. The original resolution of the egocentric view is $1280 \times 1960$ while the original resolution for the frontal view is usually $1920 \times 1080$. 
We manually take a tight crop around the face for both videos and resize the resolution to $256 \times 256$ while maintaining the aspect ratio. In all experiments and comparisons we use $7500$ frames for training, $2500$ frames for validation and the rest for testing. Tab.~1--2 in the supplemental document lists the sequences used.

\section{Egocentric-Driven Videoconferencing}
\label{sec:expressionnn}

\begin{figure*}
   \centering
   \includegraphics[width=0.95\textwidth]{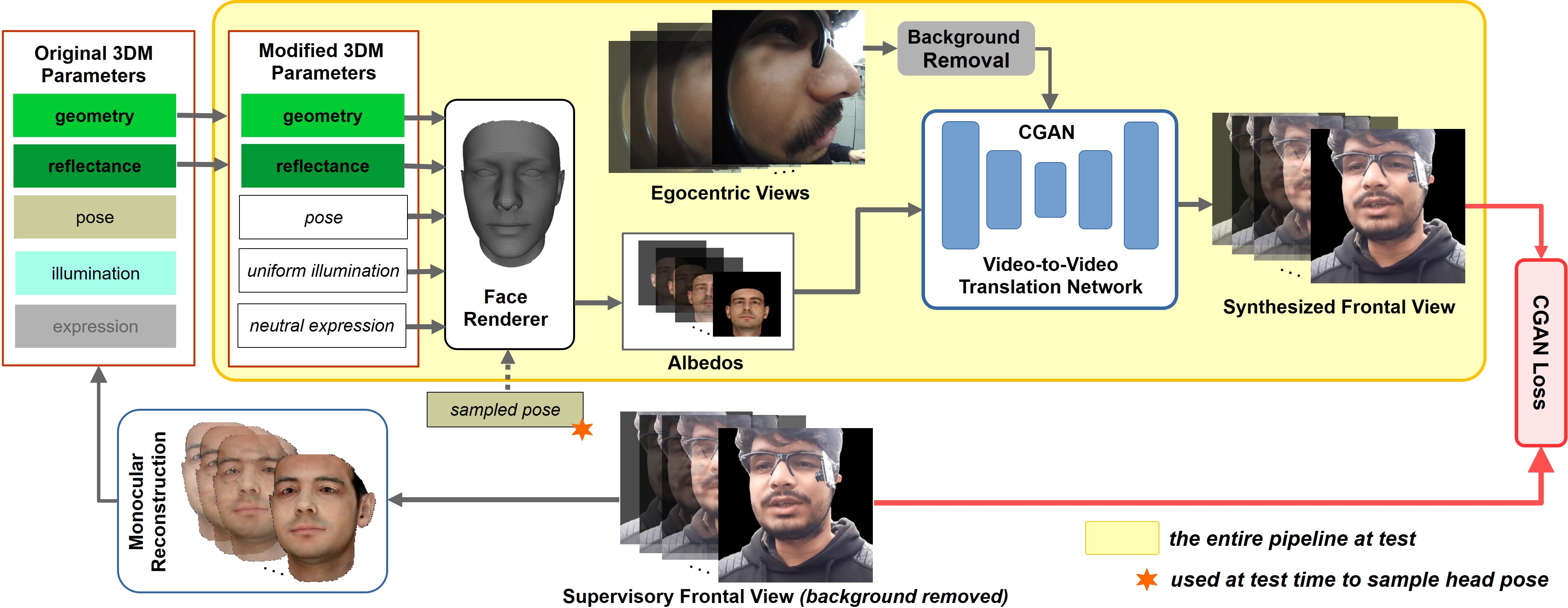}
   \caption{An overview of the proposed framework. Our approach learns to translate the egocentric view into a frontalised view. It extracts facial expressions from the egocentric view, while the head pose and identity are controlled through a parametric 3DM face model. Our approach is trained using synchronised egocentric and front view pairs. At test, results can be synthesised at arbitrarily sampled head poses (see star). {Here, the original 3DM model parameters, which were estimated during training, are modified to include no expressions, uniform illumination and the sampled head pose.}
   } 
   \label{fig:overview}
\end{figure*}

Our egocentric videoconferencing learns to photo-realistically synthesise video frames of head and upper body in frontal view based on egocentric video frames; recreating eye blinks, eye gaze, person-specific talking style and subtle non-verbal expressions, as well as the realistic appearance under various illumination. 
In Fig.~\ref{fig:overview} we show an overview of our approach.
At the core, our method is a video-to-video translation technique that gets an expression as well as a head pose conditioning as input.
The expression conditioning is purely based on the egocentric input video, and, thus, is reflecting the reality one-to-one without losing information that for example could stem from the projection onto a low dimensional 3D expression model.
As a pose conditioning, a rendering of a 3D face model with neutral expression in the target view is provided to the translation network (see Sec.~\ref{sec:pose_conditioning}).
In the following, we will describe our approach in more detail.
Our translation network is inspired by the recent success of generative models in producing photo-realistic face renderings \cite{Lombardi2019,pagan,Kim19NeuralDubbing,Fried2019,thies2019}.
We use a conditional generative adversarial network consisting of a generator network $\mathbf{G}$ and a discriminator $\mathbf{D}$.
Instead of a single image, we process a sequence of $N=11$ frames resulting in a video-to-video translation network.
The generator takes as input the egocentric views $\mathbf{E_i}$ as conditioning on the expression and renderings of the neutral face model $\mathbf{C_i}$ as pose conditioning.
It produces a sequence of photo-realistic renderings as viewed from a frontal perspective.
Note that the egocentric views not only contain expression information but also information about illumination, eye-gaze, eye-lid, tongue, etc., albeit in a possibly starkly distorted way. 
Our network is trained in a supervised manner, with paired egocentric and front-view data.
The front-view data is collected using a commodity RGB camera (see Sec.~\ref{sec:data}). 
It is used for the supervised rendering loss as well as for the extraction of the geometry, reflectance and pose that are used for the pose conditioning.
The backgrounds of both the egocentric and front-view are removed using BiSeNet~\cite{Yu18} (see Sec.~\ref{sec:background}). This allows better control of the head pose and reduces artifacts around the face borders. 

\subsection{Network Architecture}
Our generator network $\mathbf{G}$ is a U-Net-style convolutional neural network.
We stack the series of $N$ conditioning maps along the feature dimension, resulting in a input size of dimension $6N \times256\times256$.
The output of the network are $N$ RGB-frames ($3N \times256\times256$). 
The U-Net consists of~$7$~down- and up-convolutional layers with skip connections. 
All used kernel sizes have a spatial dimension of $4\times4$.
For the down-convolutions we use a stride of $2$. The up-convolutions are implemented as transposed convolutions.
The resulting U-Net based architecture contains seven levels ($128^2$,$64^2$,$32^2$,$16^2$,$8^2$,$4^2$,$2^2$) with an increasing number of feature channels per level ($64$, $128$, $256$, $512$, $512$, $512$, $512$).
The decoder mirrors in the encoder.
As a discriminator network we use a patch-based convolutional network similar to pix2pix~\cite{IsolaZZE2017}.
Instead of feeding single images to the discriminator, we input all $N$ frames of the window into the discriminator.
Thus, the discriminator works on the sequence level.
Besides the real or fake videos, the discriminator is conditioned on the stack of input images (expression and pose conditioning). 

\subsection{Training}
Our video-to-video translation network is trained according to the non-saturating game~\cite{goodfellow2014generative,IsolaZZE2017}.
The generator $\mathbf{G}$ minimizes the adversarial loss to provide outputs at a high level of video-realism, whilst the discriminator $\mathbf{D}$ maximizes the classification accuracy of real and fake videos. 
In addition to the adversarial loss, we employ a content loss and a perceptual loss: 
\begin{equation}
\label{eq:translation}
\argmin_{\mathbf{G}}~\max_{\mathbf{D}}~{E_{\textrm{A}}(\mathbf{G}, \mathbf{D}) + \lambda_1 E_{C}(\mathbf{G})} + \lambda_2 E_{P}(\mathbf{G}) \text{.}
\end{equation}
Here, $E_{\textrm{A}}(\mathbf{G}, \mathbf{D})$ is the adversarial loss, $E_{C}(\mathbf{G})$ the content loss and $E_{P}(\mathbf{G})$ the perceptual loss.
The individual losses are combined with empirically determined weights $(\lambda_1,\lambda_2)$ which are fixed to $(\lambda_1,\lambda_2)=(10.0,0.0025)$ in all our experiments.

\begin{figure*}
   \centering
   \includegraphics[width=1.0\textwidth]{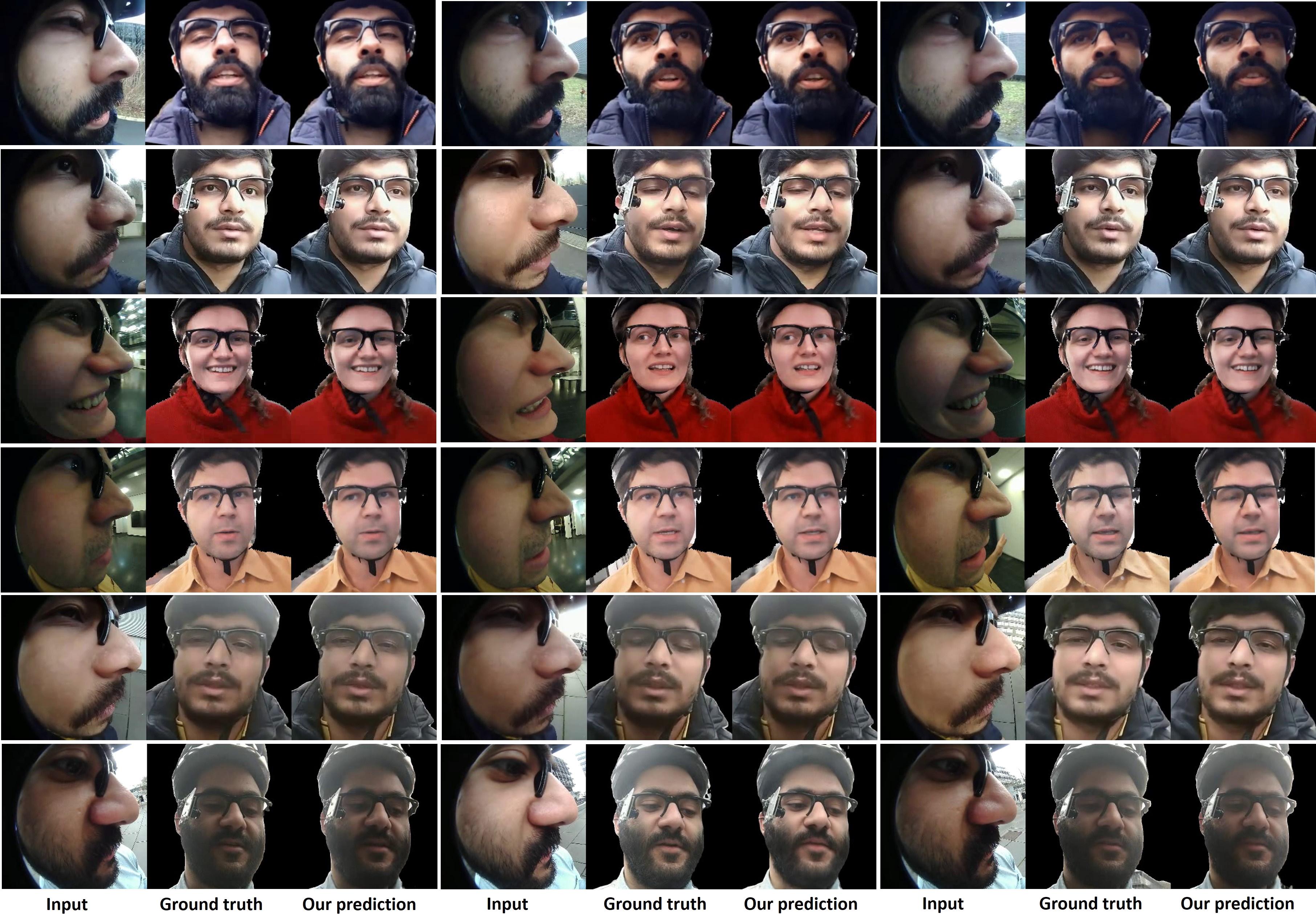}
   \caption{Our approach allows operation in a dynamic scenario. It captures a variety of facial expressions (mouth movement, eye-blinks, eye-gaze), scene illumination and produces accurate audio-lip sync.
   } 
   \label{fig:dynamic}
\end{figure*}
The adversarial loss is defined as:
\begin{equation}
E_{\textrm{A}}(\mathbf{G},\mathbf{D}) =
\mathbb{E}_{\mathbf{X},\mathbf{Y}} \big[ \!\log \mathbf{D}(\mathbf{X},\mathbf{Y}) \big] +
\mathbb{E}_{\mathbf{X}} \big[ \!\log\big(1 - \mathbf{D}(\mathbf{X},\mathbf{G}(\mathbf{X}))\big) \big] \text{.}
\label{eq:advloss}
\end{equation}
The input to the discriminator $\mathbf{D}$ is $\mathbf{X}$, and either the predicted output images $\mathbf{G}(\mathbf{X})$ or the ground truth images $\mathbf{Y}$. 
$\mathbf{X}$ are the inputs to our translation network containing the egocentric views  $\mathbf{E}$ and the pose conditionings $\mathbf{C}$.
\begin{figure*}
   \centering
   \includegraphics[width=1.0\textwidth]{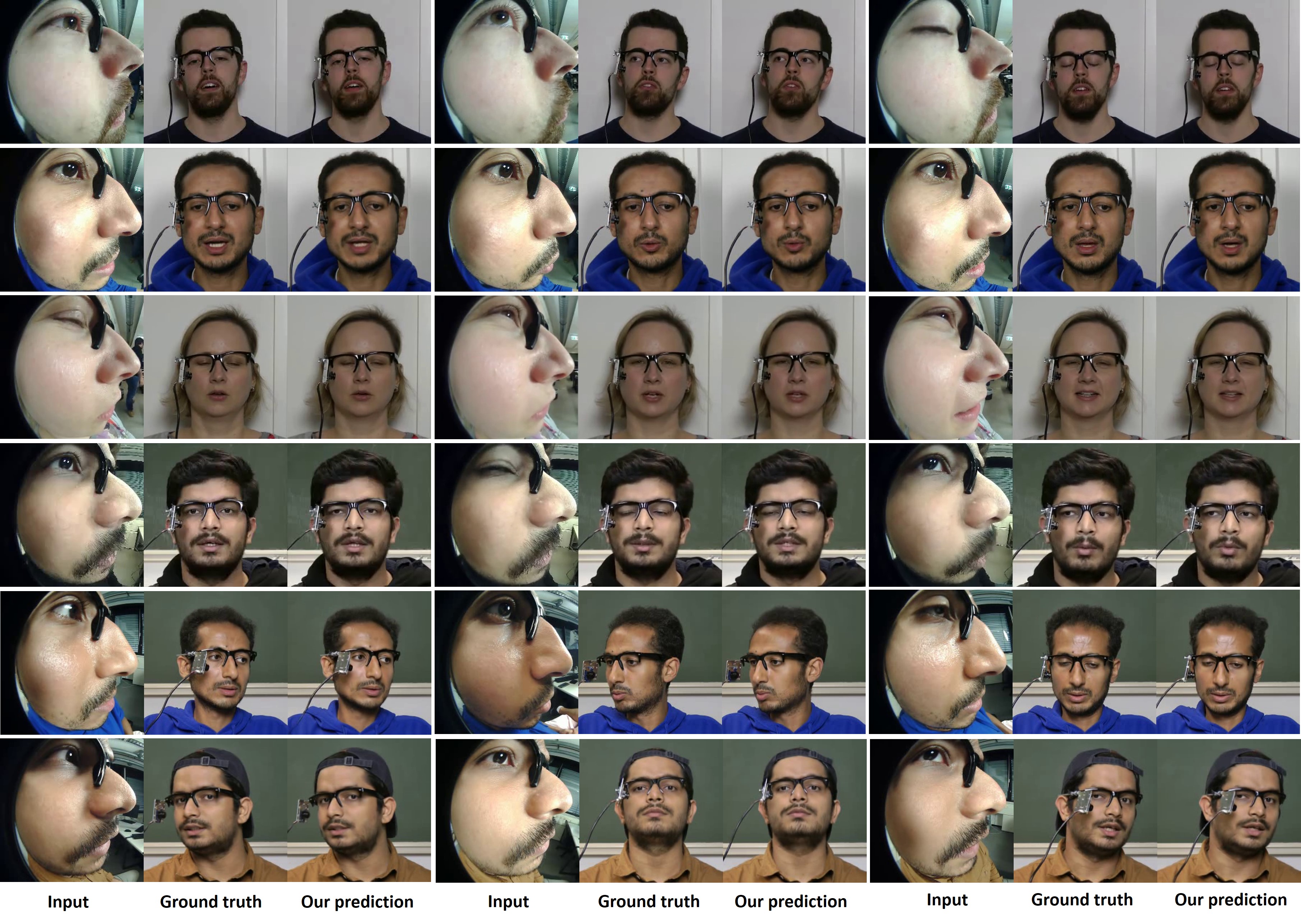} 
   \caption{Our approach also can operate in a sitting scenario. It captures mouth movements, eye-blinks, eye-gaze, head movement, different hair-styles {and a wide range of head poses (\textit{e.g.,} in the last two rows). Here, the user moves his head extremely in yaw/pitch. All sequences use $7500$ frames for training and $2500$ frames for validation.} 
   }
   \label{fig:sitting}
\end{figure*}
The $\ell_1$-based content loss enforces the output images $\mathbf{G}(\mathbf{X})$ to resemble the ground truth $\mathbf{Y}$ through
\begin{equation}
E_{C}(\mathbf{G}) =  \mathbb{E}_{\mathbf{X},\mathbf{Y}}\big[ \norm{ \mathbf{Y} - \mathbf{G}(\mathbf{X}) }_1 \big] \text{.}
\label{eq:contentloss}
\end{equation}
Finally, as a perceptual loss $E_{P}$, we employ the pretrained VGG-Face network \cite{Parkhi15}.
With respect to an $\ell_1$-norm, we measure the distance between the predicted and the ground truth images at the outputs of the convolutional layers $1, 6, 11, 18, 25$ of the VGG-Face network.
%

\begin{comment}
{Temporal Coherence:}
%
In order to improve temporal coherence, we process multiple frames at the same time, \textit{i.e.,} multiple per-frame conditionings are input to our network and the network also predicts multiple output frames.
%
The sequence is examined using a moving window of size 11 frames. 
%
Our network produces 11 frames at once, however, we only take the middle frame as the output. 
%
The loss in measured on all output frames, thus, enforcing temporal coherence.
%
Note, during test time we apply a sliding window and select the middle frame as final per frame output.
%
\end{comment}

\subsection{Pose Conditioning for Relative Head Movements}
\label{sec:pose_conditioning}
To enable the control of head movements in the target view, we provide a synthetic face rendering as conditioning to the generator network $\mathbf{G}$.
The conditioning is based on the rendering of a 3D face model with neutral expression in the desired pose.
This conditioning gives us explicit control over the head pose in the synthesised output. 
At train time the pose parameters as well as the neutral 3D face are determined by monocular face reconstruction~\cite{thies2016face}.
As input to the monocular reconstruction approach, we use the images of the front-perspective camera.
Using the tracking information, we render the albedo of the neutral face with the per frame estimated rigid pose and the identity parameters (see Fig.~\ref{fig:overview}).

\subsection{Background Removal}
\label{sec:background}

The focus of our approach lies on the reproduction of a face that reflects the captured images of the egocentric camera in a frontal view.
We do not handle the synthesis of a dynamic background.
To this end, we remove the background in our experiments using the scene segmentation technique of BiSeNeT~\cite{Yu18}. 
For the frontal views used for training, we segment each frame and set the background to black.
We also remove the background for the egocentric view. This allow us to achieve better pose control. We estimate face segmentation mask for only one frame using BiSeNeT~\cite{Yu18}, and use it for the rest of the sequence. 
We manually adjust the mask so that the mouth is visible. 
Note how our approach treats background removal of the frontal views differently from the background removal for egocentric view. The rest of the paper will keep this distinction.

\subsection{Synthesis of a Frontalised Video at Test Time} 

Our generator network gets a sliding window of egocentric views as input and produces per window a sequence of images.
In contrast to training time, we only consider the last frame that has been predicted as the output of the examined window. 
Unless stated otherwise, we sample head poses from the training set, and concatenate the corresponding 3DMM renderings to the input of our system. 
The illumination is not explicitly defined. Instead, our solution learns it from the egocentric view. 
Our approach runs in real time. It takes 29.4 ms per frame on NVIDIA Tesla V100 for a $256 \times 256$ input. 
\section{Results} 

In the following, we report the experiments that we conducted to test our pipeline. 
To see the temporal consistency of our approach, please examine the supplemental video. 
First, we show the performance of our technique subjectively on several sequences, shot in dynamic and sitting scenarios. 
The capabilities of our approach in reproducing a wide variety of facial expressions are discussed, and its ability to reenact an avatar is shown. 
To investigate the importance of each component of our method, we conducted several ablative studies (see Sec.~\ref{sec:Numerical}). 
Specifically, we numerically quantify results by estimating the photometric error between the renderings and the ground truth frontal view.
{We also investigate other aspects of our solution, including the impact of the training data size and computational complexity.}   
{All components of our techniques are evaluated in the main and additional supplemental videos (from 8:00 to 9:25 and from 11:40 to the end, respectively).} 
In Sec.~\ref{sec:comp}, we compare against related state-of-the-art approaches. We examine a wide variety of approaches, including pix2pix~\cite{IsolaZZE2017}, hypothetical advanced implementations of state-of-the-art facial reenactment techniques~\cite{kim2018DeepVideo,thies2019}, an audio-driven reenactment approach~\cite{thies2019nvp} and an unpaired image-based translation technique~\cite{Recycle-GAN}.
%%%
Finally, we discuss a user-study carried out on 44 subjects, analysing different aspects of our approach. 
%%%

%
All our sequences contain around $14000$ frames extracted at 24 frames per second (please see the table of sequences in the additional document). We use $7500$ frames for training, $2500$ for validation and the rest for testing. 
Each sequence is trained for 100 epochs, and the model producing the lowest validation error upon Eq.~\ref{eq:translation} is used. 
We use learning rate of $0.0002$, first momentum of $0.5$ and batch size of $12$. 
{Kindly note that the $7500$ frames for training and $2500$ frames for validation (less than seven minutes in total) are required per scene. 
In the dynamic scenario, we show that this duration is 
sufficient for reenacting a pre-recorded avatar of the same person (Fig.~\ref{fig:reenactement}). 
This shows that our approach can be applied in a person-specific manner in the most practical use case of egocentric videoconferencing, \textit{i.e.,} when 
moving while talking. 
}

\subsection{Subjective Evaluation} 
\label{sec:Subj} 

\begin{figure*}
   \centering
   \includegraphics[width=.8\textwidth]{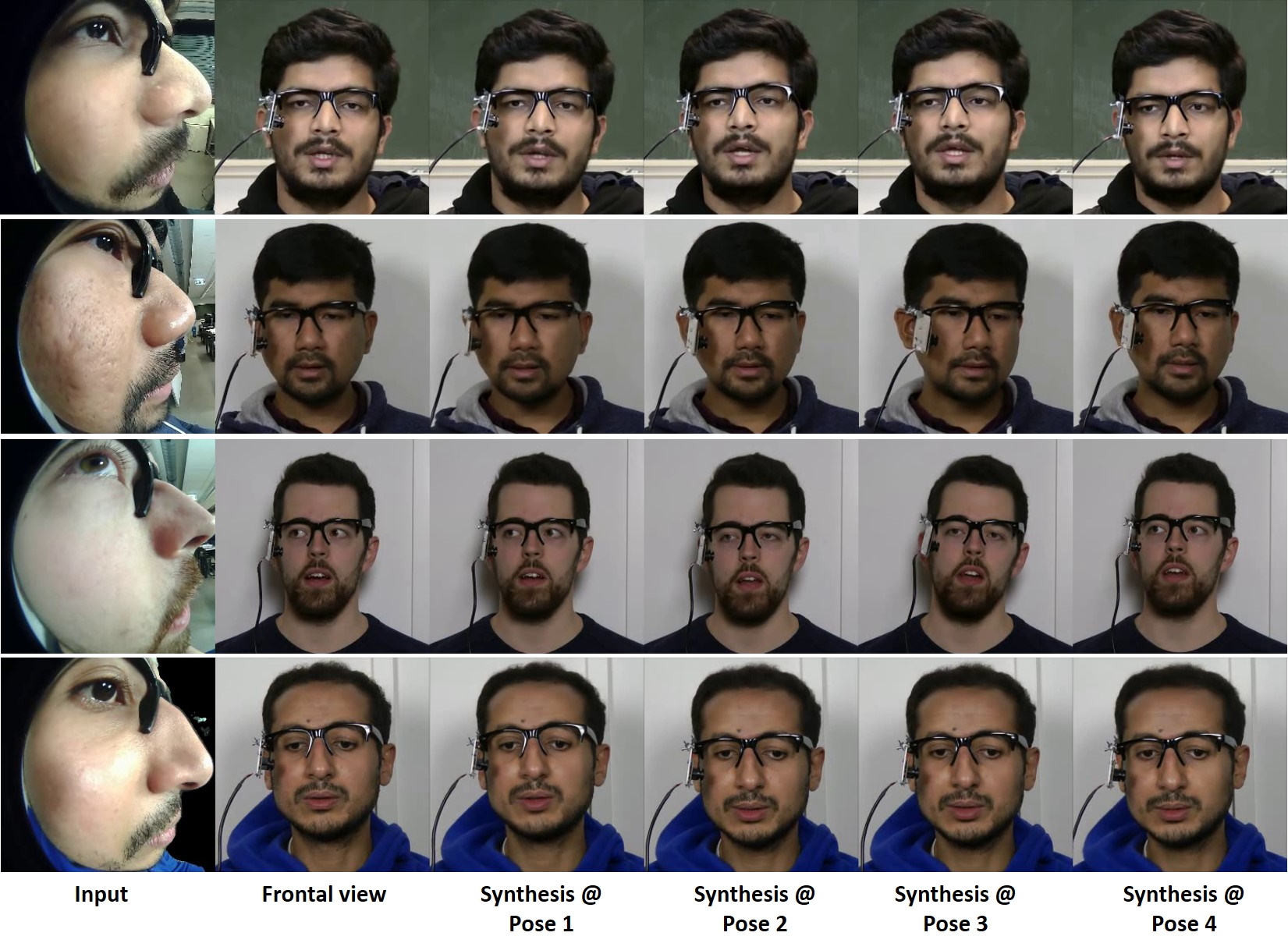}
   \caption{Our approach allows controlling the head pose during the test while maintaining the mouth movements. Here, we synthesise results at different poses sampled from the training set. Pose1 is the result synthesised at the ground truth head pose. Pose 4 is a static head pose. Please see the supplemental video.} 
   \label{fig:pose}
\end{figure*}

\begin{figure*}
	\centering
	\includegraphics[width=\linewidth]{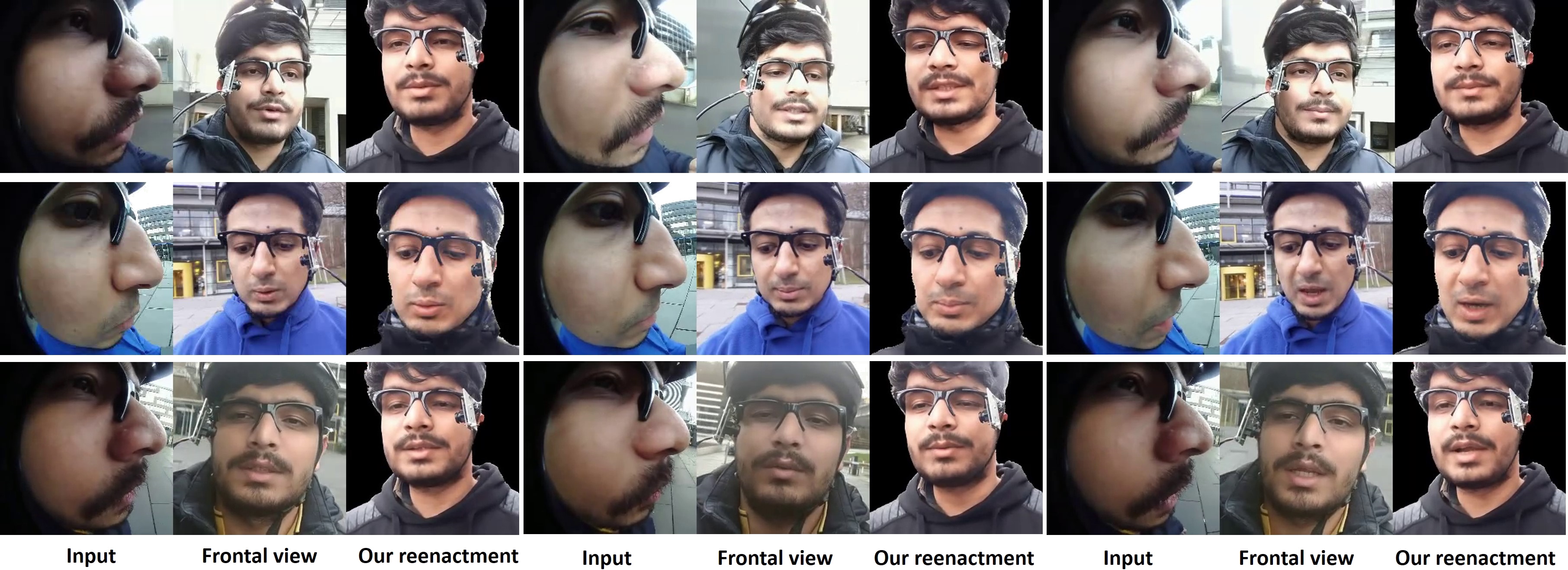}
	\caption
	{Our approach reenacts an avatar of the same person wearing different clothes and shot in a different environment, capturing mouth and eye movements.}
	\label{fig:reenactement}
\end{figure*}

Figs.~\ref{fig:dynamic}--\ref{fig:sitting} show the operation of our system in dynamic and sitting scenarios. 
Our approach reproduces mouth movements and captures eye-gaze and eye-blinks. 
It also handles subjects with heavy facial hair. 
In the dynamic scenario, our translation network learns to reproduce the scene illumination by observing the egocentric input. 
Our solution produces naturally moving head movements,  which is more evident in the sitting scenario. 
Here, results are conditioned on the ground truth head pose.  
Fig.~\ref{fig:pose} shows that results can be synthesised with different head poses. 
We randomly select a start frame from the training set and take the corresponding neutral faced 3DMM renderings as the  conditioning to our input. 
The rendered frames are taken sequentially from the start frame to ensure temporal coherency. 
Results show that we maintain mouth movement in the different poses, including a still static pose (last column). Fig.~\ref{fig:reenactement} shows that our approach can reenact avatars of the same identity. Here, we drive an avatar of the test subject using the input egocentric view and randomly sampled head pose from the training set. The avatars are wearing different clothing than the one worn at test and were recorded in a different environment. The avatars were also recorded using an egocentric camera. The clothing is considered a part of the egocentric background and hence was removed using background removal. 
This limits its interference. 
In order for reenactment to work, the egocentric camera position of the driving sequence needs to be similar to the egocentric camera position of the avatar. 
For this, we manually outline the egocentric face mask of the  avatar for just one frame. 
While wearing the device for the driving sequence, we adjust the  camera position until it overlaps with the egocentric avatar face mask. 
This adjustment is made in real time.
Our reenactments (Fig.~\ref{fig:reenactement}) are photo-realistic and temporally coherent, capturing facial expressions and eye movements. 

\begin{figure*} 
   \centering 
   \includegraphics[width=1.0\textwidth]{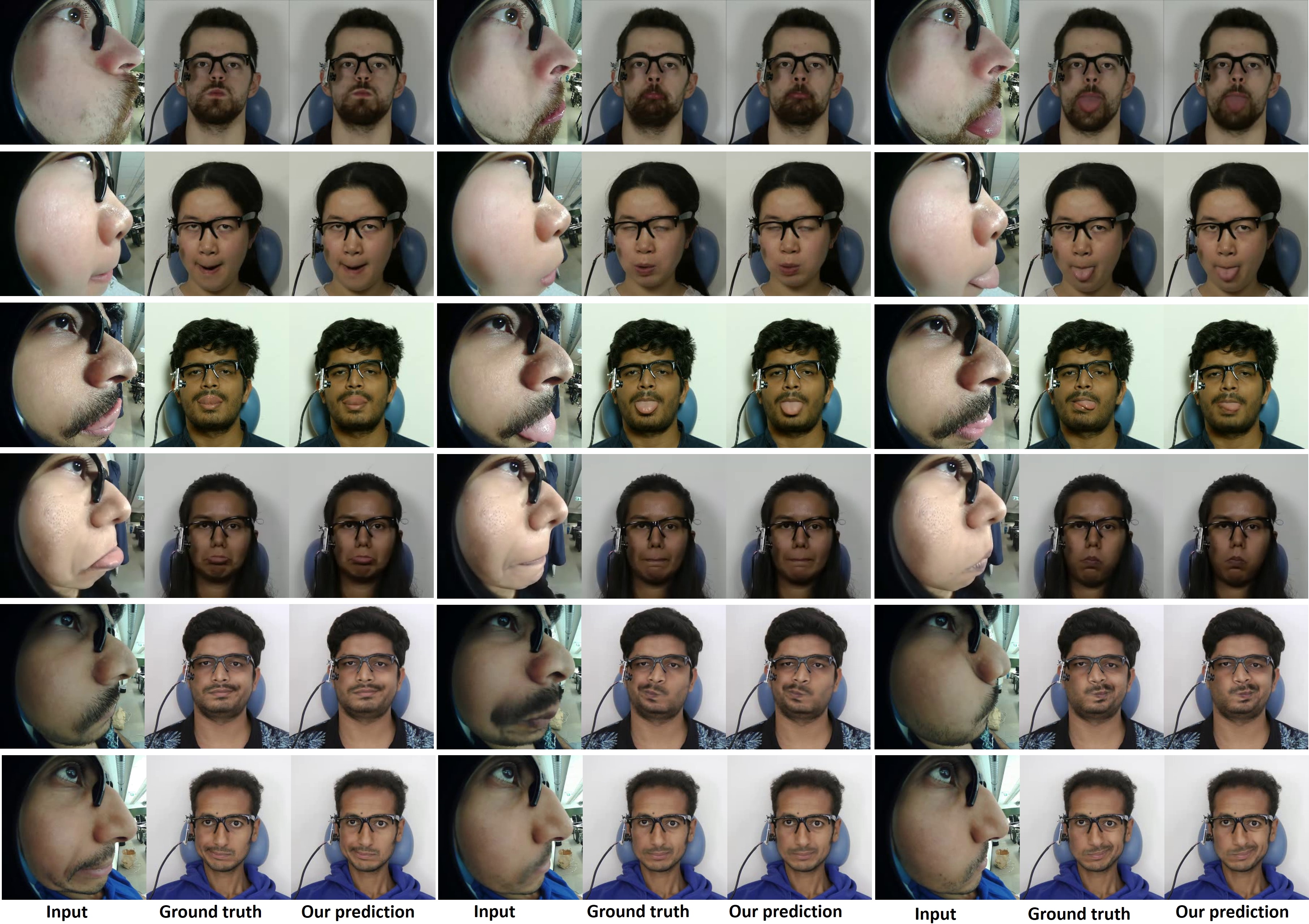} 
   \caption{Our approach captures a wide variety of expressions including bloating, lip-rolling, tongue movement and eye-blinks. These expressions are not easily captured by a 3DMM.
   {Our approach also reproduces asymmetrical smirks (see the last two rows).
   } 
   } 
   \label{fig:expressions} 
\end{figure*}

\begin{figure}[h]
	\centering
	\includegraphics[width=0.9\linewidth]{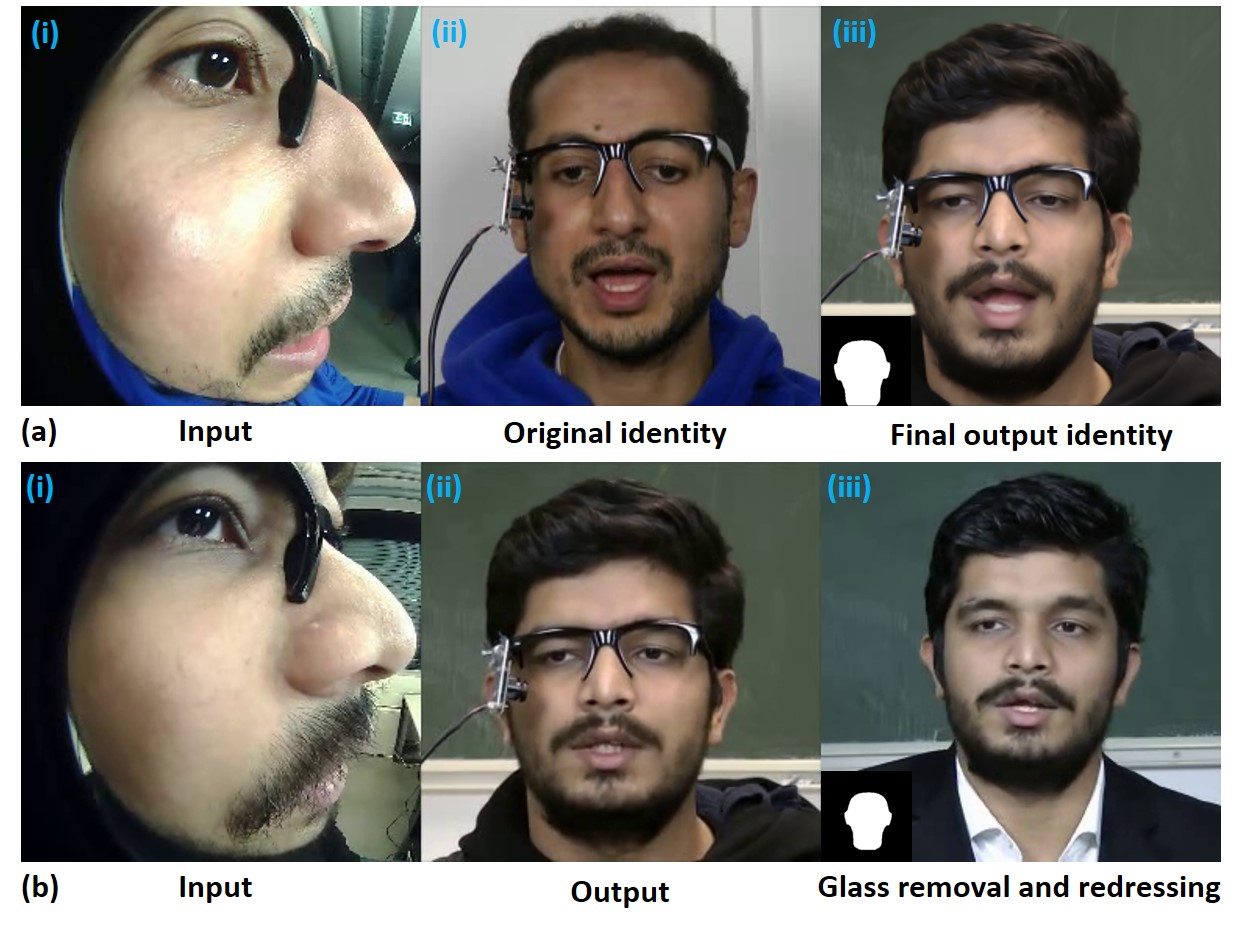}
	\caption
	{{(a) Cross-identity result where the final output identity (iii) is different from the original identity (ii). (b) An example of removing glasses and redressing the input subject while maintaining his identity. In both cases, we modify the face region of a target video (last column, inset).}}
	\label{fig:CrossID}
\end{figure}

\begin{figure}[h]
	\centering
	\includegraphics[width=0.9\linewidth]{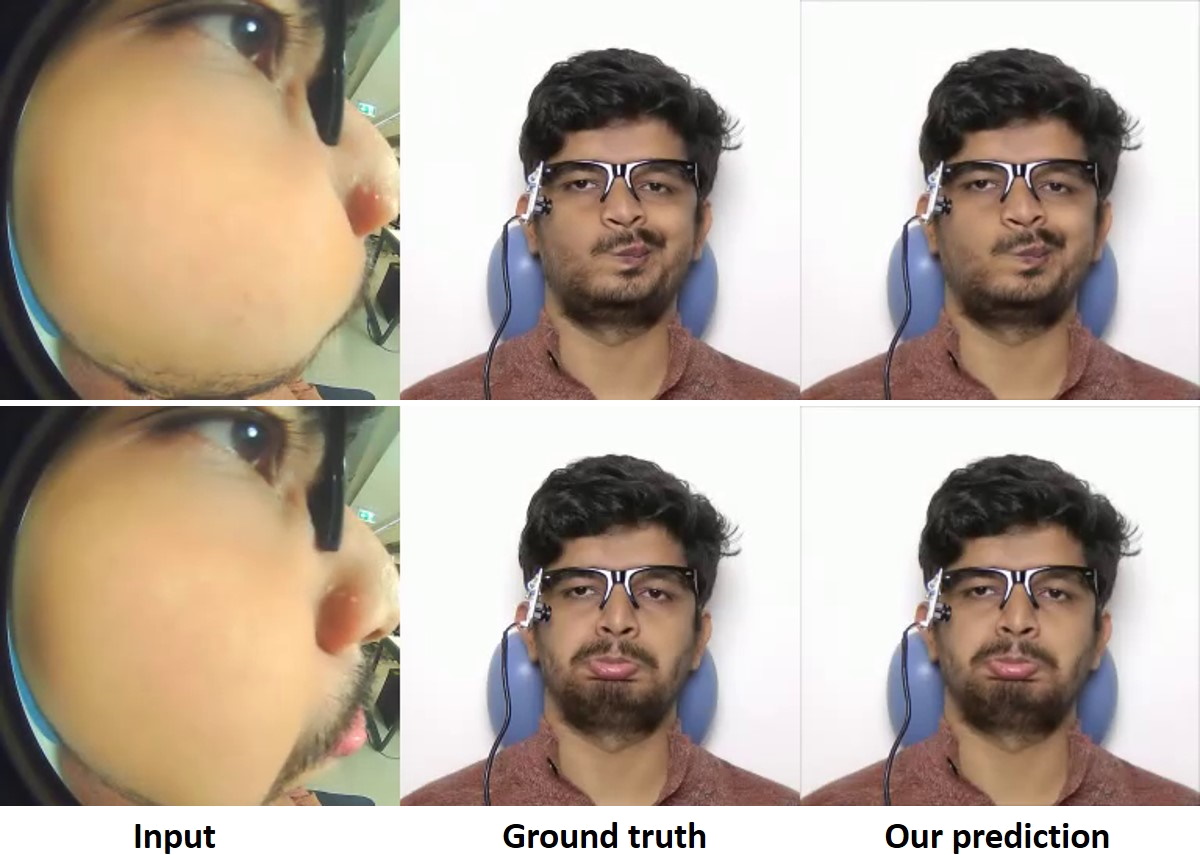}
	\caption
	{{Our approach handles extremely distorted and incomplete inputs. It produces photo-realistic results and even captures asymmetrical smirks.}}
	\label{fig:lessobstructive}
\end{figure}

In Fig.~\ref{fig:expressions}, we evaluate our approach on stress cases of expressions. Each subject was asked to repeat a set of expressions including tongue-movement, lip-rolling in, lip-rolling out and bloating. Such expressions are challenging to reproduce through parametric expression modelling. To reduce the impact of head movement on such extreme expressions, the subject's head was rested over a blue pillow. Subjects were asked to blink normally and not to move much during recording. 
Fig.~\ref{fig:expressions} shows that our technique can reproduce a wide variety of expressions with no artefacts {including asymmetrical smirks (see the last two rows).}
Results are photo-realistic and temporally coherent (please see the supplemental video). 
Note for these sequences we turned off pose conditioning as the model-fitting can struggle to disentangle head pose from the expressions.
We also did not remove the background from the egocentric view, which allowed our approach to learn and synthesise natural head movements. 
{Fig.~\ref{fig:CrossID}-a shows a cross-identity result where the final output identity (iii) is different from the original identity (ii). We first apply our frontalisation approach to the egocentric view (i), which produces (ii). We post-process (ii) using the neural renderer of ~\cite{kim2018DeepVideo}. We train the renderer on the face region only (see iii, inset) and copy the background from the target video. This dedicates more network capacity to the face which helps in better rendering the mouth interior. The same approach can remove the glasses and redress the input while maintaining his identity, see Fig.~\ref{fig:CrossID}-b. Note that results might experience some audio mis-sync. Future work can examine a dedicated loss for the mouth region~\cite{Fried2019}.
}
{Fig.~\ref{fig:lessobstructive} shows that our technique can handle less obstructing camera positions, \textit{i.e.,} the camera is moved even closer to the face. Despite the input view is more distorted and more incomplete than earlier examples, our approach still produces temporally consistent photo-realistic results and is even capable of reproducing asymmetrical smirks. As in all sequences, we used $7500$ and $2500$ frames for training and validation,  respectively. 
}

\subsection{Quantitative Evaluations}
\label{sec:Numerical}

\begin{figure*}
   \centering
   \includegraphics[width=0.9\linewidth]{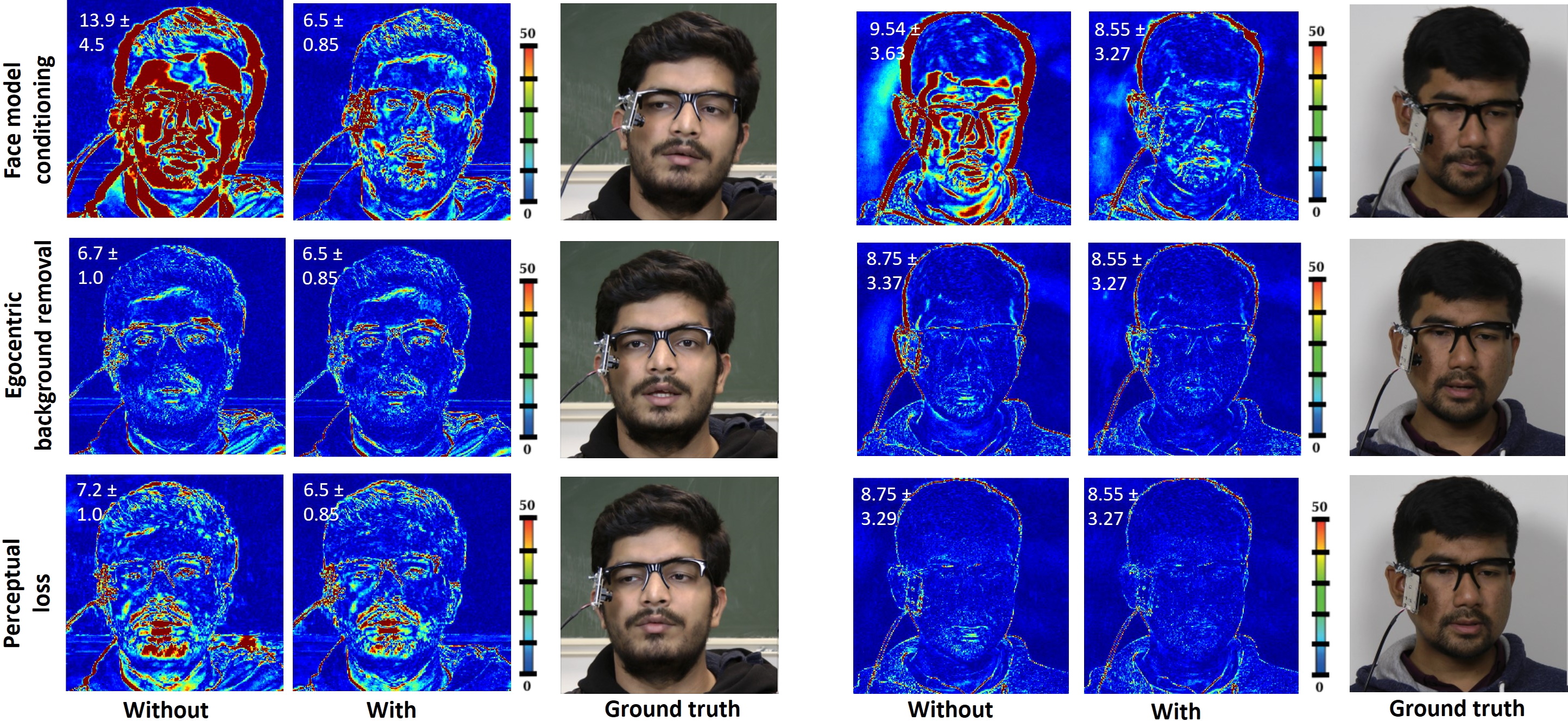}
   \caption{An ablative study of our approach. For each design choice (row) we show the photometric error ($\ell_2$ in RGB space) between our results and the ground truth. 
   Removing a specific design choice leads to more temporally inconsistent results (see std. deviation). Especially, removing the face model conditioning leads to significant artefacts. The right subject has a larger standard deviation of the error than the left subject as his original motion pattern is more dynamic.} 
   \label{fig:ablation}
\end{figure*}

\begin{figure}[h]
	\centering
	\includegraphics[width=1.0\linewidth]{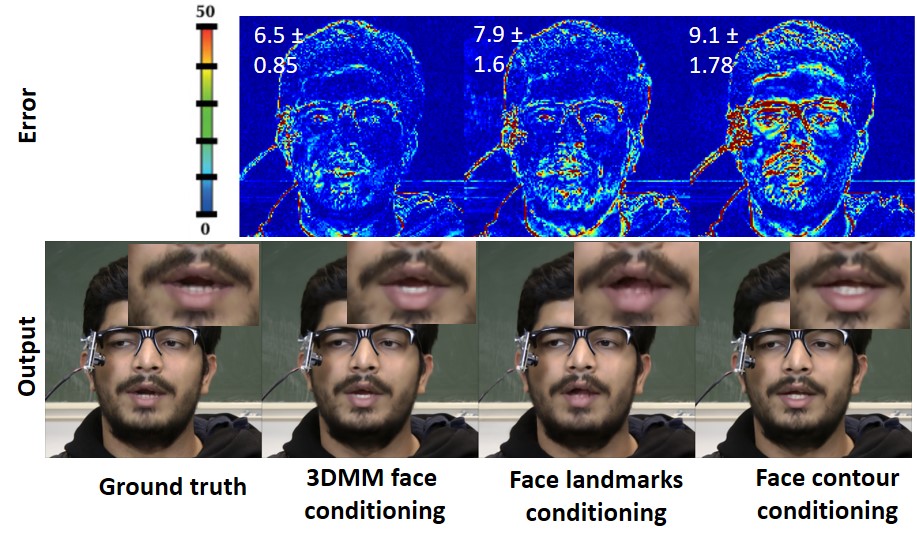}
	\caption
	{{The effect of using face contours and face landmarks~\cite{bulat2017far} for pose conditioning as opposed to using a 3DM face model. Face landmarks produce localised artefacts as in the mouth interior (see zoomed regions). Face contours produce spatially spread artefacts (see error maps) which flicker temporally. 3DMM based conditioning, however, produces the best results.}}
	\label{fig:facemodeling}
\end{figure}

\begin{figure}[h]
	\centering
	\includegraphics[width=0.9\linewidth]{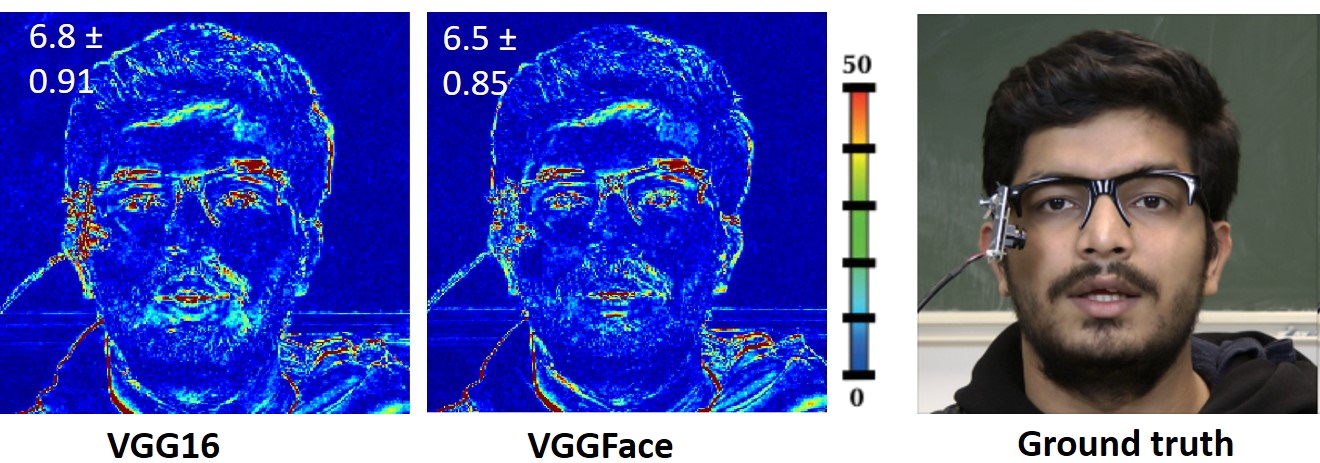}
	\caption
	{A VGG-Face~\cite{Parkhi15} perceptual loss produces more temporally coherent results than a VGG16~\cite{Simonyan15} loss. This is reflected here by the lower error mean and standard deviation.}
	\label{fig:vgg16}
\end{figure}

\begin{figure}
	\centering
	\includegraphics[width=0.9\linewidth]{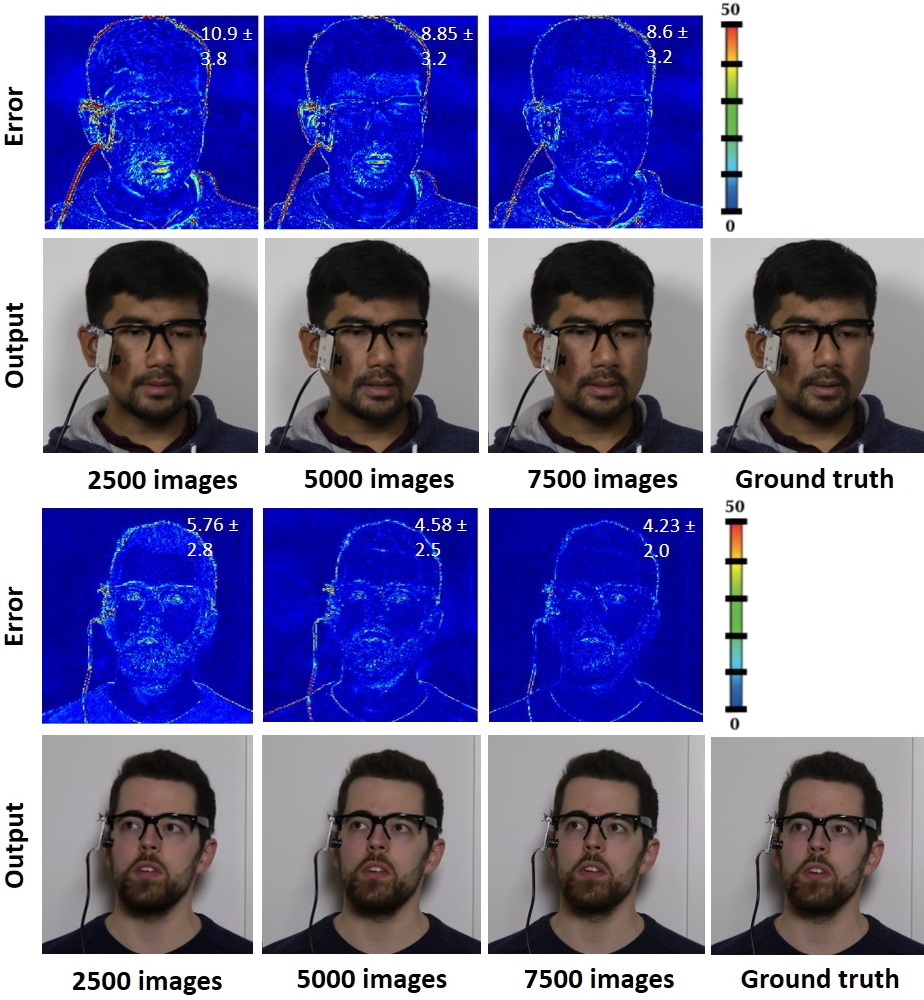}
	\caption
	{Impact of different training data sizes. Here, we show qualitative results and the photometric error visualisation over the full sequence. Results show that we can produce similar results with variable training sizes, while the qualitative difference is barely noticeable (especially in $7500$ vs $5000$).}
	\label{fig:trainsize}
\end{figure}

\begin{figure}
	\centering
	\includegraphics[width=.8\linewidth]{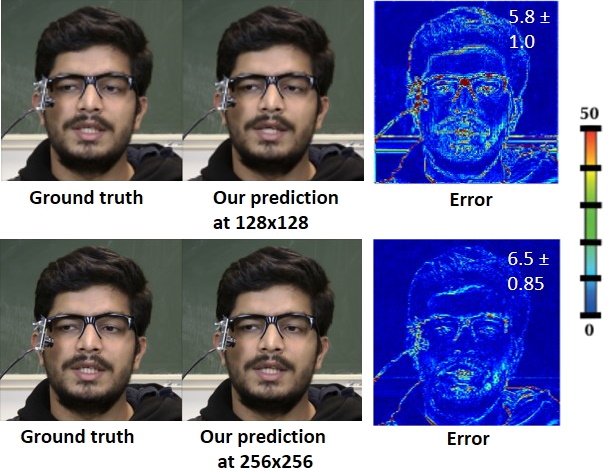}
	\caption
	{Our approach achieves similar reconstruction error when processed at $128 \times 128$ and $256 \times 256$. Hence while $128 \times 128$ leads to more speedup, the visual quality loss is not significant (see the additional video).}
	\label{fig:realtime}
\end{figure}

We performed multiple experiments to assess the importance of each design choice of our pipeline (Figs.~\ref{fig:ablation}--\ref{fig:vgg16}). 
Most artefacts are in the form of temporal shakiness as well as  unnatural head movements and deformations and therefore best viewed in a video (please see our supplemental video). 
To aid in examining the results on paper, we performed quantitative evaluations by estimating the photometric error between the output and the ground truth frontal view. 
The error is estimated as the Euclidean distance on the RGB space (in the range of 0-255). We report the average per frame error and the corresponding standard deviation. 
This is always shown at the top (right or left corner) of the error heat-maps.
The standard deviation gives a good indication to the degree of temporal stability; higher variance corresponds to stronger temporal shakiness. 
Note that for numerical evaluations, results must be synthesised at the ground truth pose, and hence we do so when necessary (Figs.~\ref{fig:ablation}--\ref{fig:realtime}, \ref{fig:pix2pix}). 

Fig.~\ref{fig:ablation} shows that removing pose conditioning leads to significant artefacts. This is due to the one-to-many expression-to-frontal view mappings. 
Conditioning on the pose without removing the egocentric background leads to video artefacts. 
Background removal of the egocentric view helps in disentangling head pose from the rest of the facial components, and thus allows better synthesises, especially at different head poses (see Fig.~\ref{fig:pose}). 
This is an essential feature of our solution as it reduces the reliance on knowing the ground truth pose during the test. 
Removing perceptual loss leads to shaky results and artefacts. 
This is shown by the higher error.
Fig.~\ref{fig:vgg16} shows that a VGG-Face~\cite{Parkhi15} perceptual loss produces more accurate results than a VGG16~\cite{Simonyan15} loss. 
Fig.~\ref{fig:facemodeling} investigates different face representations for pose conditioning. Conditioning using the facial landmarks instead of the 3DMM-based renderings produces unstable results. 
{This is due to the sparse nature of facial landmarks  which loses many elements of the face structure. In addition, facial landmarks contain an expression component, and hence it is challenging to disentangle and control the head pose.} 
While a face contour representation can contain a weaker expression component and is popular for conditional face generation~\cite{neuraltalkingheads}, it still produces artefacts in the form of temporal flickering (see error map). 
3DMM-based pose conditioning, however, produces the best visual and numerical results. 
We hypothesise this is due to its ability to disentangle expressions from the identity and pose. 

{Fig.~\ref{fig:trainsize} shows an ablative study with respect to the training data size. We trained models using $7500$, $5000$ and $2500$ frames. 
For each model, we report the photometric error between the ground truth and our results, as well as the corresponding means and standard deviations. 
We also examine the visual quality and compare it against ground truth. 
Results show no significant loss in quality between training with $7500$ frames and $5000$ frames, as indicated numerically and visually (see additional supplemental video). 
This shows the potential of our approach in using less training data.
However, some temporal flickering can occur due to less training data, especially when training with $2500$ frames.}
Fig.~\ref{fig:realtime} examines the processing speed of our approach on two input resolutions. On Tesla v100, one frame of a $256 \times 256$ resolution takes 29.4 ms to process, while a $128 \times 128$ frame takes 23.75 ms. On Titan 2080 Ti, one frame takes 45.1 ms for $256 \times 256$ and 27.6 ms for $128 \times 128$. 
While the speedup gain in processing $128 \times 128$ images is noticeable, Fig.~\ref{fig:realtime} shows that the loss in visual quality may not be significant (please see the additional supplemental video). 
We argue that $128 \times 128$ is suitable for our videoconferencing application, especially if the final rendered output is to be viewed on a mobile phone screen. 
The processing time is taken as the average of three sequences (each processed five times). 
It also includes image reading and writing times to simulate possible delay due to video transmission. 
Note for real-time processing, the per-frame processing speed should be at most 40 ms (25 fps). 

\subsection{Comparisons to Related Techniques}
\label{sec:comp}

We compare our approach against related image- and audio-based techniques. 
In all our comparisons, we use the same training, validation and test set used by our technique: 
\begin{itemize}
    \item pix2pix~\cite{IsolaZZE2017} is a paired image-to-image translation approach using a conditional GAN. We train pix2pix to learn the frontal mapping from the egocentric view. 
    \item We implemented hypothetical egocentric-compatible versions of two high-quality reenactment techniques~\cite{kim2018DeepVideo,thies2019}. The aim of this study is to investigate the limitations inherited from an expression model-based solution. We examine both Deep Video Portraits ~\cite{kim2018DeepVideo} and Deferred Neural  Rendering~\cite{thies2019}. Since these approaches can not handle our egocentric view, we instead used the corresponding frontal view as input.
    \item Neural Voice Puppetry (NVP)~\cite{thies2019nvp} is a recent audio-driven reenactment approach. It takes as input the audio signal and modifies the lower-part of the face of a pre-recorded video.
    \item We compared our reenactment against state-of-the-art unpaired image translation technique, Recycle-GAN~\cite{Recycle-GAN}. This is to assess its ability to reenact avatars wearing clothes and shot in environments different from the driving sequence. 
\end{itemize}

\begin{figure}
	\centering
	\includegraphics[width=0.8\linewidth]{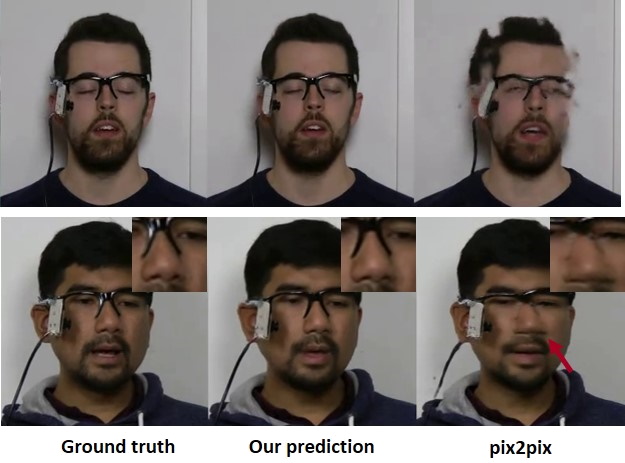}
	\caption
	{pix2pix~\cite{IsolaZZE2017} generate temporally inconsistent results with visual artefacts (see the first row and the red arrow/nose).}
	\label{fig:pix2pix_visual}
\end{figure}

\begin{figure}
	\centering
	\includegraphics[width=0.8\linewidth]{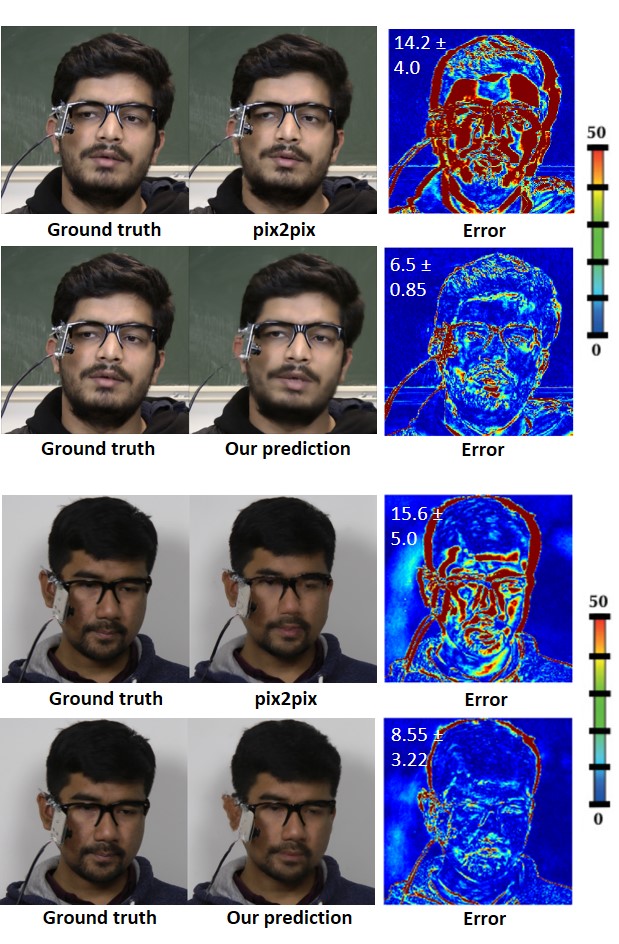}
	\caption
	{Quantitative evaluations show that our approach significantly outperforms pix2pix~\cite{IsolaZZE2017}.} 
	\label{fig:pix2pix}
\end{figure}

\begin{figure*}
   \centering
   \includegraphics[width=1.0\textwidth]{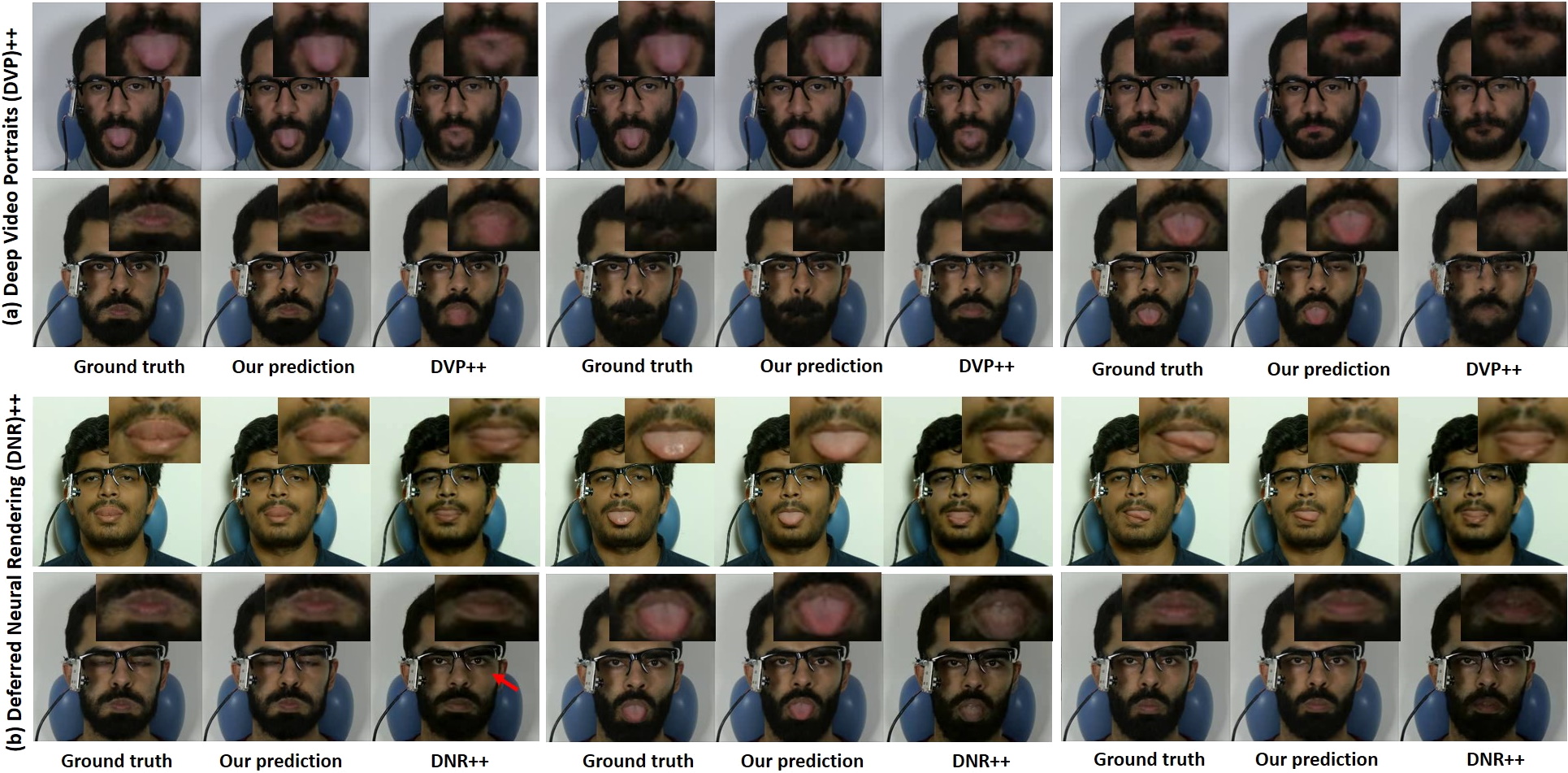}
   \caption{Comparison against the advanced (hypothetical) implementation of Deep Video Portraits~\cite{kim2018DeepVideo} and Deferred Neural Rendering~\cite{thies2019}. Here, the face model parameters are estimated from the supervising front-view camera. Despite that, our approach still produces significantly better results (see mouth zoom-ons). It better handles mouth and tongue movements, eye-blinking (see red arrow) and heavy bearded subjects.} 
   \label{fig:modelBasedpp}
\end{figure*}

\begin{figure*}
	\centering
	\includegraphics[width=1.0\linewidth]{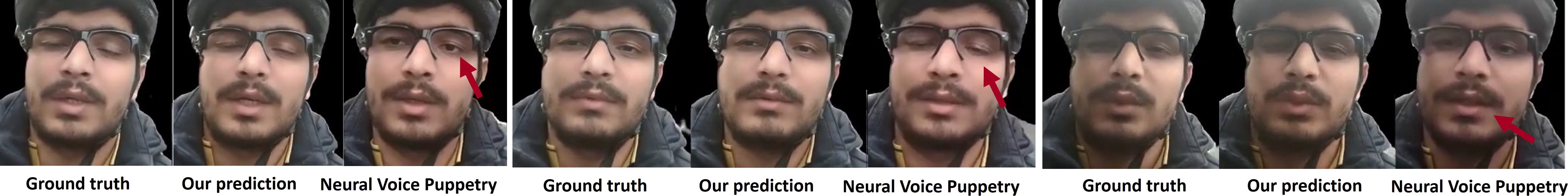}
	\caption
	{Neural voice puppetry~\cite{thies2019nvp} is an audio-driven reenactment solution and hence does not capture ground truth non-verbal expressions (see eye blinks in the first two columns, red arrows) nor scene illumination. It is also sensitive to background audio noise which could lead to inaccurate mouth movements (third column, red arrow). Our approach resembles ground truth more accurately.
	}
	\label{fig:nvp}
\end{figure*}

\begin{figure*}
   \centering
   \includegraphics[width=0.8\textwidth]{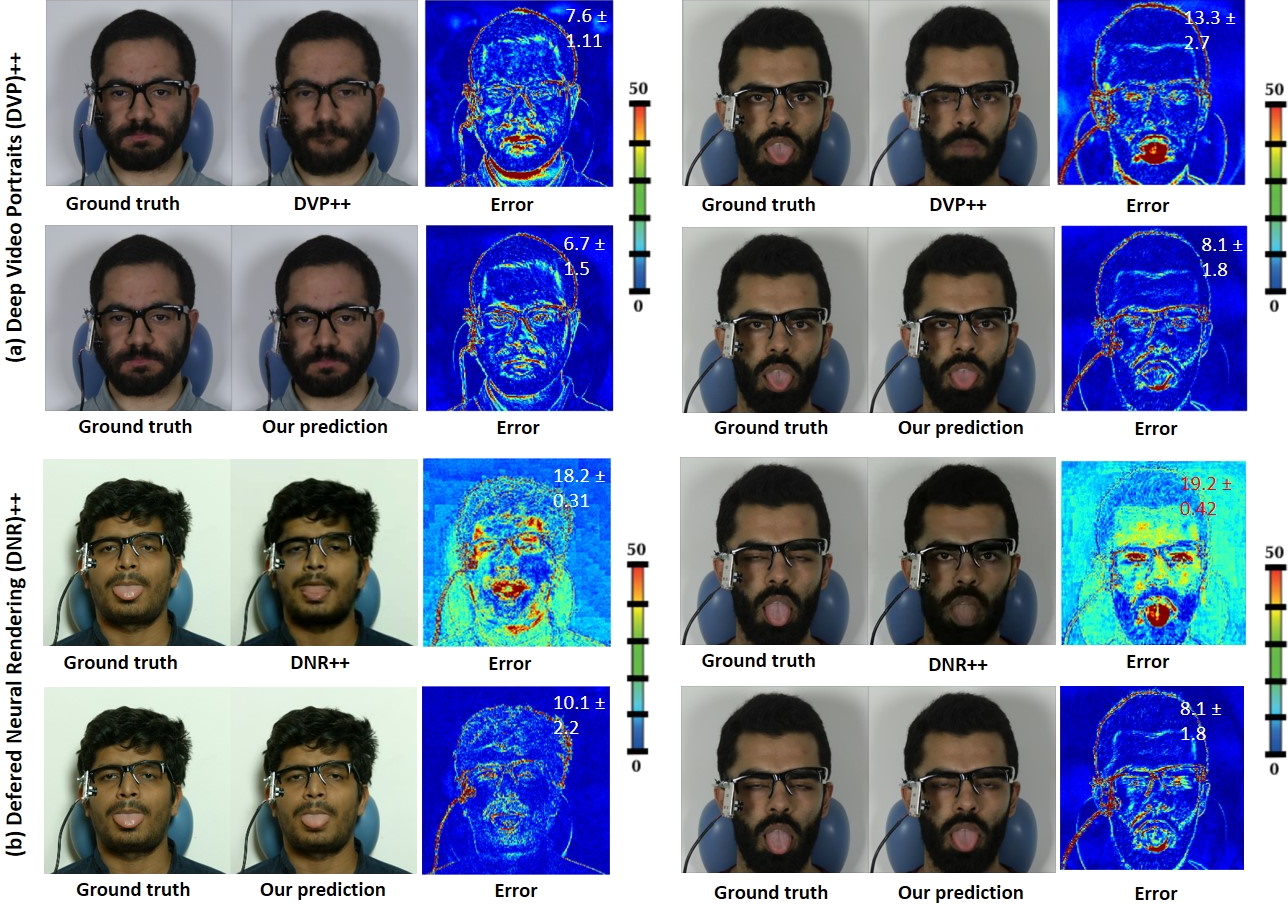}
   \caption{Our approach achieves lower photometric error in comparison to an advanced implementation of Deep Video Portraits~\cite{kim2018DeepVideo} (top) and  
   an advanced implementation of Deferred Neural Rendering~\cite{thies2019} (bottom). The lower standard deviation of Deferred Neural Rendering++ is due to output with less motion than ground truth.} 
   \label{fig:dvpdnrnum}
\end{figure*}

\begin{figure}
	\centering
	\includegraphics[width=0.85\linewidth]{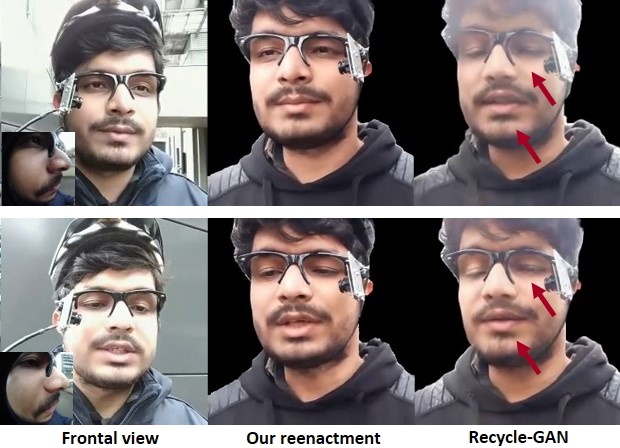}
	\caption
	{Our approach reenacts an avatar using expressions from the egocentric input (shown in insets). While our reenactment captures the expressions of the frontal view (left), Recycle-GAN struggles with reproducing mouth and eye movements (see red arrows) and leads to strong artefacts (see video).  
	}
	\label{fig:recycleGAN}
\end{figure}

Subjective and numerical evaluations show that pix2pix struggle with handling our egocentric views and generate noticeable artefacts (see Figs.~\ref{fig:pix2pix_visual} and \ref{fig:pix2pix}). Results are often shaky with unnatural movements as reflected by the high error variance (please see supplemental video). 
The expression-model used in Deep Video Portraits~\cite{kim2018DeepVideo} and Deferred Neural  Rendering~\cite{thies2019} limits the range of expressions that can be reproduced. In addition,  fitting the face model to a heavily bearded subject can be erroneous and produce significant visual artefacts (see Fig.~\ref{fig:modelBasedpp}, mouth region). 
Such poor performance is also reflected quantitatively in 
Fig.~\ref{fig:dvpdnrnum}.
Neural Voice Puppetry~\cite{thies2019nvp} produces photo-realistic results (see Fig.~\ref{fig:nvp}). However, since it modifies the lower part of a prerecorded video, it does not reproduce the ground truth upper face movement (\textit{e.g.,} blinking) nor captures the scene illumination. Furthermore, as being an audio-based solution, it is sensitive to background audio noise. This could lead to inaccurate mouth movements (Fig.~\ref{fig:nvp}, third row). 
Finally, reenactment using the state-of-the-art unpaired image translation technique Recycle-GAN~\cite{Recycle-GAN} fails in reproducing eye blinks and mouth movements (see Fig.~\ref{fig:recycleGAN}, red arrows). This leads to noticeable artefacts, with the avatar not speaking (see the video). Our approach nevertheless can drive a target avatar using egocentric expressions, even when wearing clothes not seen during training. 

\begin{figure}
	\centering
	\includegraphics[width=0.85\linewidth]{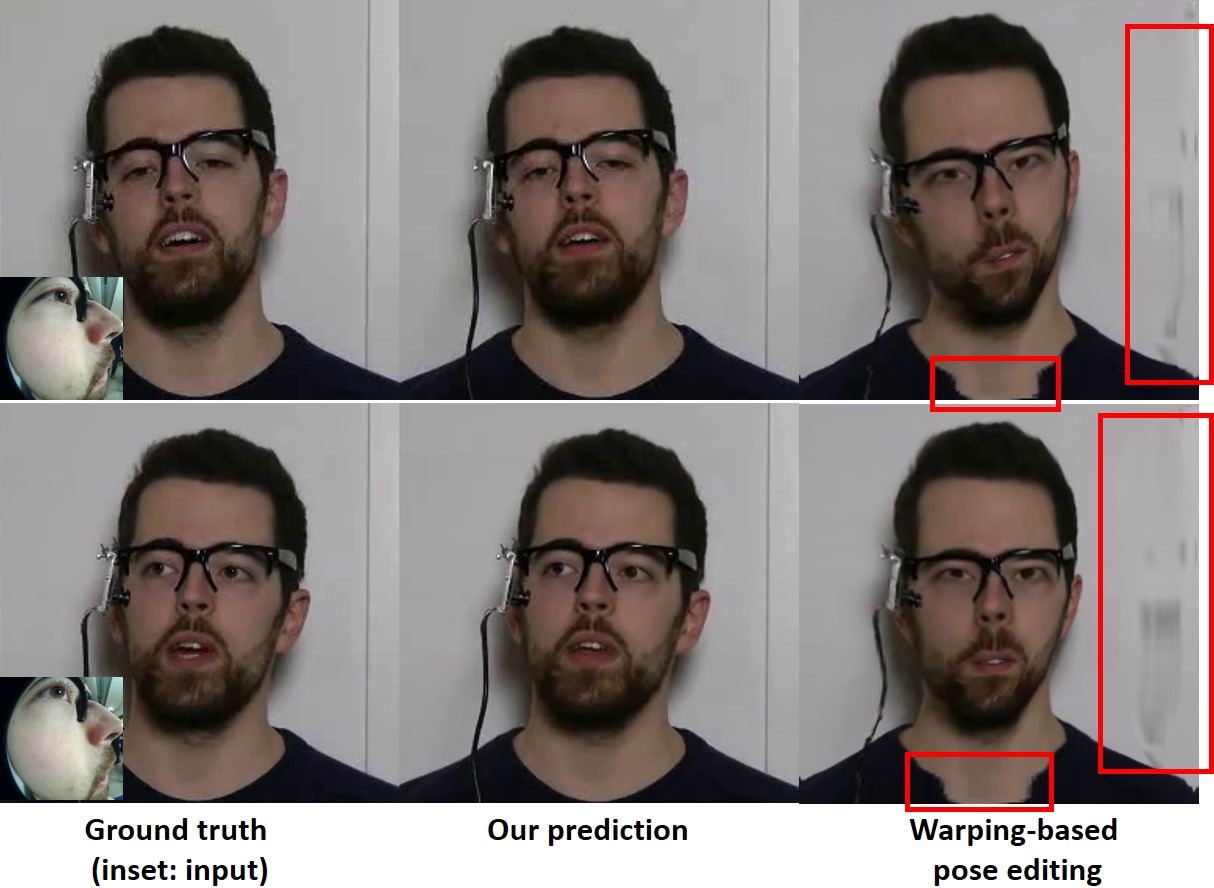}
	\caption
	{{Comparing our predictions against warping-based head pose synthesis. Here, the target head pose is estimated using~\cite{Ruiz18} and X2Face~\cite{Wiles18} drives a single source image by editing its pose. Such an approach generates significant artefacts in the head geometry, as well as in the background and body torso (see red regions).
	}}
	\label{fig:x2face}
\end{figure}

{We also compared our method against a warping-based head pose synthesis approach based on X2face~\cite{Wiles18}.
X2face~\cite{Wiles18} is used to edit the head pose according to the yaw, pitch and roll of a target video.
We assume the ideal case of perfect pose-free frontalisation and edit the head pose of a single source image.
We estimate the head pose of the target video using the approach of Ruiz~\etal~\shortcite{Ruiz18}.
In Fig.~\ref{fig:x2face}, we show that such a warping-based approach leads to severe artefacts in the background and also torso region (see the red regions).
In contrast, our approach outputs photo-realistic results while also not only synthesising the head pose changes but also the facial expressions. 
}

\subsection{User Study} 
We performed two surveys to assess the results produced by our approach visually. 
In the first survey, we compared our approach against pix2pix~\cite{IsolaZZE2017} and examined the importance of using our face model conditioning. 
We showed the users two long and continuous videos for pix2pix, and two long and continuous sequences for our solution without face model conditioning. Each sequence is $45$ seconds long and contains to its side (either left or right) the output video of our full solution. 
The order of the videos was shuffled. 
A participant was asked the following question: "which video looks more natural to you", and his/her answer was reported by either choosing left or right. 
Users were asked to listen to the audio. Out of $41$ participants, $84\%$ rated our results more natural than pix2pix. 
In addition, our approach was rated $85.2\%$ more natural than when no conditioning is used.
These results show the significance of our approach and its design choices. 
It also shows that when no conditioning is used, results similar to pix2pix are produced, (with difference of $1.2\%$, less than 1 participant). 
In the second survey, we investigated how real our results look. 
We displayed twenty videos, ten being real and ten produced by our approach. 
The videos ordering was shuffled. 
Participants were asked to rate their agreement to the statement "this video looks natural to me" using a linear scale from strongly agree, agree, do not know, disagree and strongly disagree. 
All backgrounds in the videos were black. 
We asked users to ignore this and any artefacts occurring around them in their ratings. 
We also asked the users to listen to the audio. 
Out of 44 participants, $62.7\%$ agreed that our results look natural. 
Note that $77.5\%$ of the respondents agreed that the original videos look natural. 
While this shows a $22.5\%$ baseline error due to users suspicion, it also shows that our approach produces decent and naturally-looking results. 

\begin{figure}
	\centering
	\includegraphics[width=0.9\linewidth]{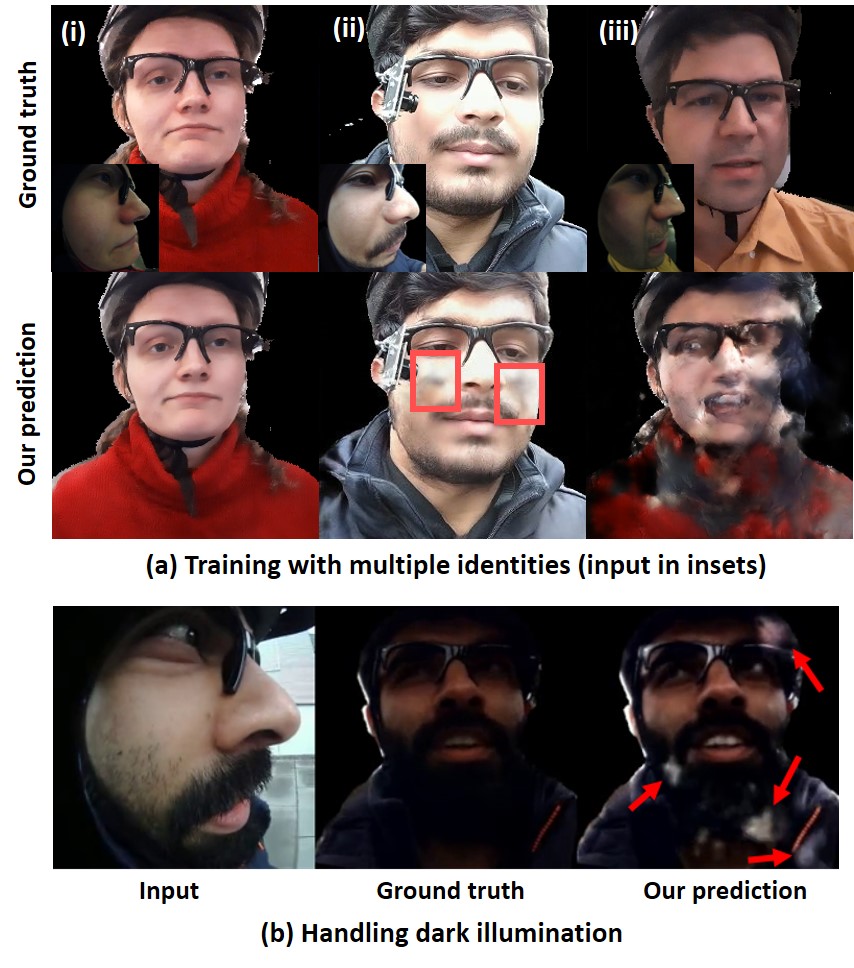}
	\caption
	{{(a) We examined training a model on four different identities. Our technique reproduces the correct identity if it is included in the training (i \& ii). However, results can experience temporal flickering (see red regions, ii). Testing on unseen identities produces incorrect renderings (iii). Nevertheless, the method tries to reconstruct the most similar training identity, that is of (i).} (b) Our approach struggles in handling sequences with very dark illumination and can produce white-spots like artefacts (see red arrows).} 
	\label{fig:limitations} 
\end{figure} 
\section{Limitations and Future Work}

In our experiments, we processed a wide variety of sequences in dynamic and sitting scenarios.
At the moment, our solution is person-specific and constrained to the expressions seen at training time.
{Fig.~\ref{fig:limitations} (a) summarizes the result of mixing four different identities in one model. Our technique reproduces the correct identity if it is included in the training set (see Fig.~\ref{fig:limitations}-a, i \& ii).
Results, however, can experience temporal flickering (see Fig.~\ref{fig:limitations}-a, ii, red regions). Testing on unseen identities hallucinates incorrect renderings with strong artefacts (see Fig.~\ref{fig:limitations}-a, iii). Here, the network attempts to reproduce the training identity that looks most similar to the test identity.
A future research direction for addressing these issues could be to expand the network capacity to accommodate for the variations that occur over multiple identities.} 
Our work focuses on reconstructing faces, and hence future efforts can investigate rendering backgrounds in a dynamic scenario.
While our technique can synthesise results at arbitrary head poses (shown in Fig.~\ref{fig:pose} and the supplemental videos), synthesising results at the ground truth head pose can aid photo-realism. 
Future work could investigate learning the ground truth head pose from the audio signal~\cite{shlizerman2017audio,ginosar2019gestures} or directly obtaining it from an IMU as in ~\cite{li2015facial,olszewski2016high}.
Finally, our technique can struggle with scenes shot in very dark illuminations, leading to artefacts (see Fig.~\ref{fig:limitations}-b).

\section{Conclusion}
We introduced the first real-time hands-free egocentric videoconferencing approach for mobile eyewear devices. 
Our technique takes as input distorted and incomplete egocentric facial views. 
It learns frontal facial expressions, and the coarse facial details such as the identity and head pose from a parametric 3D head model. 
We achieve state-of-the-art frontalisation results that are temporally stable and expressive. 
The experiments show that our approach operates well in dynamic and sitting scenarios, and reenacts avatars of the same person wearing different clothes. 
It also captures a wide variety of challenging expressions such as tongue and depth inducing movements, not easily captured by an expression model-based solution.
We compared against a variety of related techniques and evaluated the design choices of our solution subjectively and numerically. 
We believe that our approach is a stepping stone towards new hybrid facial animation systems that can use both a parametric model as well as image-translation based conditioning for capturing fine details. 

\begin{acks}
This work was funded by the ERC Consolidator Grant 4DRepLy (770784). 
We also acknowledge support from Technicolor.
We especially thank Ankita Chanda Roy for her help with data recording. We also would like to thank Jalees Nehvi, Gereon Fox, Mallikarjun B R, Ikhsanul Habibie, Vikramjit Sidhu, Moritz Kappel, Franziska Mueller, Varshini Muthukumar, Lingjie Liu, Edgar Tretschk and Alexandra Theobalt. For help with running experiments, we thank Jalees Nehvi and Gereon Fox. 
\end{acks}

\bibliographystyle{ACM-Reference-Format}
\bibliography{main}

\end{document}